\documentclass[preprint,authoryear,12pt]{elsarticle}
\makeatletter
\def\@xfootnote[#1]{%
  \protected@xdef\@thefnmark{#1}%
  \@footnotemark\@footnotetext}
\makeatother
\usepackage{graphics}
\usepackage{multirow}
\usepackage{amssymb}
\usepackage{lscape}
\usepackage{color}

\journal{Icarus}

\begin{document}

\begin{frontmatter}

\title{Is the Eureka cluster a collisional family of Mars Trojan asteroids?}

\author[addr1]{Apostolos A.~Christou\corref{cor1}}
\ead{aac@arm.ac.uk}
\ead{Fax: +44 2837 527174}
\author[addr1]{Galin Borisov}
\author[addr2]{Aldo Dell'Oro}
\author[addr3]{Alberto Cellino}
\author[addr1]{Stefano Bagnulo}
\address[addr1]{Armagh Observatory, College Hill,
           Armagh BT61 9DG, Northern Ireland, UK}
\address[addr2]{Osservatorio Astrofisico di Arcetri,
			 Largo Enrico Fermi 5, 
			 I-50125 Florence,
			Italy}
\address[addr3]{Osservatorio Astronomico di Torino,
			Via Osservatorio 20,
			Pino Torinese,
			10025 Torino,
			Italy}           
			
\cortext[cor1]{Corresponding author}

\begin{abstract}
We explore the hypothesis that the Eureka family of sub-km asteroids in the $\mbox{L}_{5}$ region of Mars could have formed in a collision. 
We estimate the size distribution index from available information on family members; model the orbital dispersion of collisional fragments; and carry out a formal calculation of the 
collisional lifetime as a function of size.
We find that, as initially conjectured by \cite{Rivkin.et.al2003}, the collisional lifetime of objects the size of (5261) Eureka is at least a few Gyr, significantly longer than for similar-sized Main Belt asteroids. In contrast, the observed degree of orbital compactness is inconsistent with all but the least energetic family-forming collisions. Therefore, the family asteroids may be ejecta from a cratering event sometime in the past $\sim1$ Gyr if the orbits are gradually dispersed by gravitational diffusion and the Yarkovsky effect \citep{Cuk.et.al2015}. The comparable sizes of the largest family members require either negligible target strength or a particular impact geometry under this scenario \citep{Durda.et.al2007,Benavidez.et.al2012}. Alternatively, the family may have formed by a series of YORP-induced fission events \citep{Pravec.et.al2010}. The shallow size distribution of the family is similar to that of small MBAs \citep{Gladman.et.al2009} interpreted as due to the dominance of this mechanism for Eureka-family-sized asteroids \citep{Jacobson.et.al2014}. However, our population index estimate is likely a lower limit due to the small available number of family asteroids and observational incompleteness. Future searches for fainter family members, further observational characterisation of the known Trojans' physical properties as well as orbital and rotational evolution modelling will help distinguish between different formation models.
\end{abstract}
   

\end{frontmatter}


\section{Introduction}
Trojan asteroids are objects confined by solar and planetary gravity to orbit the Sun $60^\circ$ ahead, or behind, a planet's position 
along its orbit \citep[see eg][]{MurrayDermott1999}. Trojans of Jupiter, Neptune and Mars are stable over the age of the solar system, 
dating from its formation. The taxonomies of Jupiter Trojans are indicative of primitive, geologically unprocessed bodies \citep{Grav.et.al2012}.
By contrast, the much smaller population of Martian Trojans exhibits a wide range of taxonomies, suggesting diverse origins \citep{Rivkin.et.al2003,Rivkin.et.al2007}. 

As capture of asteroids into a stable Trojan configuration with Mars is implausible in the present solar system \citep{SchwarzDvorak2012}, these objects were 
likely deposited at their present locations when the solar system had not yet reached its final configuration (A.~Morbidelli, in \citet{Scholl.et.al2005}). 
\citet{Rivkin.et.al2003} proposed that the collisional lifetime of Mars Trojans is longer than asteroids in the main belt and, therefore, that the objects we observe today 
are near their original sizes. Recently, \citet{Christou2013} identified a compact orbital cluster of Martian Trojans (the ``Eureka family'' after its largest member, 5261 Eureka) containing most of the known population, including newly-identified Trojans. In the same work, \citeauthor{Christou2013} argued that a collision plausibly formed 
this cluster but it could also have been produced by YORP-induced rotational fission \citep{Pravec.et.al2010}. 
In \citet{Cuk.et.al2015}, the orbits of a compact group of Trojan test particles ejected from Eureka were propagated in time under planetary gravitational perturbations and the Yarkovsky effect. Those authors found that the group is likely a genetic family formed roughly in the last Gyr of the solar system's history. Here, we focus on
 the problem of the family's formation. Specifically, we address the question of whether the cluster could have formed by collisional fragmentation of a progenitor body.
Our approach is three-pronged: Firstly, we exploit a recent increase in the size of the known population and perform an appraisal of the size distribution of observed family members as a clue to the formation mechanism.  Then, we apply models of collisional fragmentation, seeking to reproduce the observed orbital distribution of family members with plausible kinematical properties of the family formation event. Finally, we perform a rigorous calculation of the collisional lifetime of Martian Trojans to determine if, and when, collisional disruption is likely to have occurred.

This paper is organised as follows: In the following Section, we review the currently available information on Martian Trojans and the Eureka family. In Section 3 we look at the size distribution of family members confirmed to-date. In Section 4 we construct a probabilistic model of the collisional dispersion and escape of Trojan fragments and apply it to the case at hand. In Section 5 we perform a calculation of the lifetime of objects in the Martian Trojan clouds against collisional disruption.
Finally, in Section 6 we carry out a synthesis of results from the different lines of investigation and present our conclusions in Section 7.

\section{The observed population of Martian Trojans} 
Although Mars plays host to a population of co-orbital objects \citep{Connors.et.al2005}, here we are concerned with {\it stable} Trojans, in other words asteroids that have librated around the
$\mbox{L}_{4}$ and $\mbox{L}_{5}$ Lagrangian equilibrium points of Mars for an appreciable fraction of the age of the solar system. \citet{Scholl.et.al2005} showed by numerical integration that there are at least three such objects: (5261) Eureka \& (101429) 1998 $\mbox{VF}_{31}$ at $\mbox{L}_{5}$ and (121514) 1999 $\mbox{UJ}_{7}$ at $\mbox{L}_{4}$. 
New simulations by \citet{Christou2013} and by \citet{deLaFuenteMarcoses2013} added a further four: (311999) 2007 $\mbox{NS}_{2}$, (385250) 2001 $\mbox{DH}_{47}$, 2011 $\mbox{UN}_{63}$ and 2011 $\mbox{SC}_{191}$. Finally, \citet{Cuk.et.al2015} reported two additional objects, 2011 $\mbox{SL}_{25}$ and 2011 $\mbox{UB}_{256}$, recovered during the 2013/14 opposition and consequently confirmed as stable Trojans by integrating their orbits. Table \ref{tab:eureka family} lists their orbital elements. The quantity $D$ refers to the half-amplitude of libration of the mean longitude $\lambda$ around $\mbox{L}_{5}$. Remarkably, all 6 asteroids identified as Trojans in 2013 and 2015 are Eureka family members.

Compositional information on the three brightest Trojans shows that Eureka belongs to the rare, olivine-dominated, A-type taxonomic class while 1998 $\mbox{VF}_{31}$ shows a spectrum dominated by pyroxene and not directly related to Eureka \citep{Rivkin.et.al2003,Rivkin.et.al2007}. 1999 $\mbox{UJ}_{7}$'s featureless spectrum \citep{Rivkin.et.al2003} and low albedo \cite[$p_{\rm v}=0.048\pm0.012$;][]{Mainzer.et.al2012} suggest a ÒprimitiveÓ taxonomy, typical of bodies in the Outer Main Belt and Jupiter Trojan clouds \citep{Grav.et.al2012}. Assuming albedos of 0.2, 0.23 \& 0.06 for the respective classes within the Bus-DeMeo taxonomy \citep[5261: A class; 101429: S class; 121514: C/D class;][]{DeMeo.et.al2009,DeMeoCarry2013} yields, through the formula  
$$
D= \frac{1329}{\sqrt{p_{\rm v}}}10^{-H/5}\mbox{,}
$$
diameters of 1.8, 1.1 \& 2.2 km respectively. Therefore the three brightest Trojans are similar in size, 1-2 km across.
Analysis of observations during year one of the WISE reactivation mission gave $D=1.88$$\pm$$0.23$ km, $p_{\rm v}=0.18$$\pm$$0.05$ for Eureka \citep{Nugent.et.al2015}, consistent with our above estimates of size \& albedo for that asteroid and within the range of albedo values obtained by WISE \citep{Maziero.et.al2011} and IRAS \citep{Tedesco.et.al2004} for other so-called monominerallic olivine asteroids \citep{Sanchez.et.al2014}.

Recently, it was confirmed that asteroids (311999) 2007 NS$_2$ and (385250) 2001 DH$_{47}$, the second and third largest members of the Eureka family, share the same spectral reflectance properties as (5261) Eureka. In particular, all three objects display features that are diagnostic of an olivine-dominated surface composition \citep{Borisov.et.al2016,Polishook.et.al2016}. Since the implied taxonomy is uncommon, it further strengthens the case for a common origin of the members of this group. In other words, we have here a real \emph{family} of objects sharing a common parent body, rather than a cluster of unrelated asteroids. 

\section{Size distribution of the family}
\label{sec:size}

\begin{figure}
\vspace{-3cm}
\centering
\includegraphics[width=150mm,angle=0]{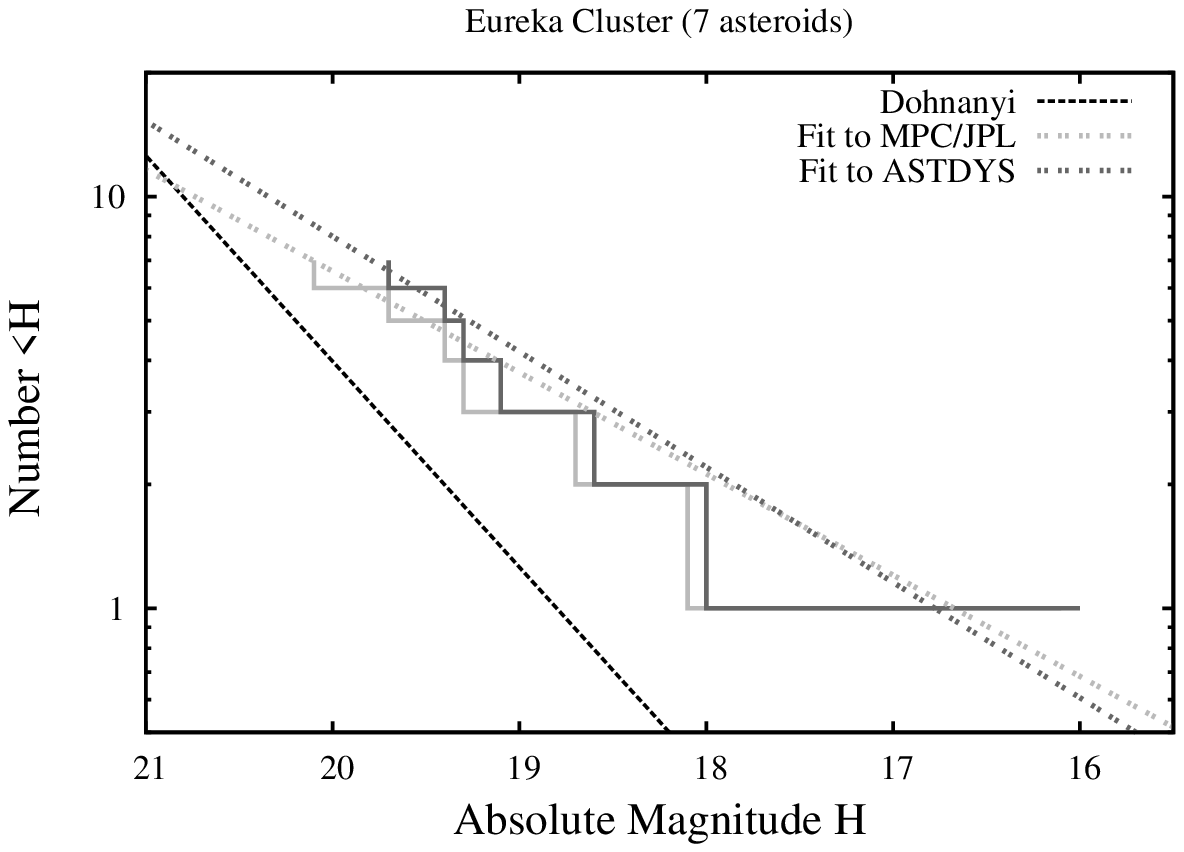}
\caption[Cumulative distribution of absolute magnitudes for Eureka family asteroids, using data from MPC/JPL (light grey line) and ASTDYS (dark grey line). The straight double-dotted lines represent power law fits to the respective distributions. The black dashed line indicates the slope expected for a collisionally-evolved population according to \citet{Dohnanyi1969}.]{}
\label{fig:population_fits}
\end{figure}

The small sample size notwithstanding, the size distribution of family members is diagnostic of the formation mechanism 
 \citep{Tanga.et.al1999,Jacobson.et.al2014}.
 Assuming that the size distribution of the objects has the form
\begin{equation}  \label{eq:size_distr}
N(>s) = A\mbox{ }s^{\hspace{0.2em}-\alpha}
\end{equation}
where $s$ is the asteroid diameter and $\alpha$ and $A$ are constants, it is simple to show that the corresponding distribution of absolute magnitudes is
\begin{equation} \label{eq:h_distr}
N(<H) = B\mbox{ }10^{\hspace{0.2em}\beta H}
\end{equation}
where $\alpha = 5  \beta$ and $B$ is related to $A$, $\alpha$ and the albedo. In this case, the magnitude distribution appears as a line in the plane $\log \left(N(<H)\right)$ vs $H$. 
A nonlinear least-squares fit utilising the Marquardt-Levenberg algorithm with inverse-variance weights yields $\beta$$=$0.25$\pm$0.02 \& 0.28$\pm$0.03 for the JPL/MPC and AstDys data respectively or, equivalently, 1.25 \& 1.4 for N($>$D). Note that AstDys $H$ values for the fainter objects are systematically higher than JPL's. MPC magnitudes tend to be underestimated for the fainter objects \citep{Juric.et.al2002}. In any case, the slopes are considerably shallower than the $\beta=0.5$ expected for a  population of collisional fragments \cite[][Fig.~\ref{fig:population_fits}]{Dohnanyi1969}. Interestingly, \citet{Gladman.et.al2009} found an exponent of $\sim 0.3$ for the distribution of asteroids with $15< H\lesssim 18$ in the Main Belt, similar to what is found here for the Trojans. This was interpreted by \citet{Jacobson.et.al2014} as being due to the action of YORP-induced rotational fission modifying the size distribution at the small end of the population ($D \lesssim 6$ km or $H \gtrsim 15$). A caveat lies in our premise, stated in the following Section, that the completeness limit of the Trojans is the same as that of NEOs, where 50\% completeness is achieved at $H=19.0-19.5$ according to \citet{HarrisDAbramo2015}. The faintest Trojans fall into or near this range, so this may have skewed the observed distribution towards a shallower slope. Incidentally, extrapolation of our power law fits implies a significant number of fainter, as yet undiscovered, Eureka cluster members, 13-22 up to $H=22$ and 57-99 up to $H=24$. These numbers should be considered as lower limits if, as pointed out above, our number statistics of observed Trojans are biased by observational completeness. This part of the population, if it exists, should be discoverable by future observational surveys \citep[e.g.~the Large Synoptic Survey Telescope;][]{Jones.et.al2015,Todd.et.al2012b}. In the near-term, the mapping now in progress by the {\sl Gaia} satellite may help confirm candidate Trojans as well as discover new objects with sizes near the present completeness limit \citep{Todd.et.al2014}.  

Finally, for our slope estimates we find an upper limit of $r=0.9$ for the mass ratio between the largest remnant and the parent body, similar to that obtained ($r=0.875$) if we sum up the volumes of the known family members. This is a useful limiting case to consider in the ensuing calculations and also allows to estimate the minimum size of the Eureka family parent body as $1.88/ \left(0.9\right)^{1/3} \simeq$ 2 km.

\section{Collisional dispersion of Trojans}
\label{sec:dispersion}

\subsection{Theoretical framework}
\label{sec:theory}
To a first approximation, a collisional fragment of a Trojan must satisfy the condition
 \begin{equation} \label{eq:esc_crit}
 |\Delta a + a_{\rm PB} - a_{\rm Mars} | < a_{\rm crit} 
 \end{equation}
 to remain a Trojan,
 where $a_{\rm PB}$ and $a_{\rm Mars}$ are the osculating semimajor axes of
 the parent body and Mars respectively. The quantity $a_{\rm crit}$ corresponds to the widest
 possible libration amplitude for a Trojan, equal to 
 \begin{equation} \label{eq:a_crit}
 a_{\rm crit} = \sqrt{8 \mu /3} \mbox{ } a_{\rm Mars}\mbox{,}
 \end{equation}
 
$\mu$ representing the Mars-Sun mass ratio and $\Delta a$ the change in semimajor axis caused
 by the ejection velocity of the fragment. This is related to the orbital elements of the parent body at the time
 of ejection through the expression \citep{MurrayDermott1999}
  \begin{equation} \label{eq:a_dv}
 \Delta a = 2 a^{2}/h \left( \Delta {\rm v}_{r} e sin f  + \Delta {\rm v}_{\theta} \left( 1 + e \cos f \right) \right)
 \end{equation}
 
 where $\Delta {\rm v}_{r}$ and $\Delta {\rm v}_{\theta}$ are two components of the ejection velocity vector in a body-centred reference frame $(\hat{{\rm r}}=\vec{{\rm r}}/|\vec{{\rm r}}| \mbox{, }\hat{\theta}= \vec{{\rm r}} \times \vec{h} /  |\vec{{\rm r}} \times \vec{h}|\mbox{, }\hat{h}= \vec{{\rm r}} \times \vec{{\rm v}} /  |\vec{{\rm r}} \times \vec{{\rm v}}| )$ defined through the instantaneous heliocentric state vector ($\vec{{\rm r}}$, $\vec{{\rm v}}$) of the parent asteroid. 
 
We note that Trojan collisional fragments may also escape if their orbital eccentricity and/or inclination exceed the boundary of the stable region of width $\sim 0.15$$\times$$15^{\circ}$ identified by \citet{Scholl.et.al2005}. However, Gauss' equations imply that, for a given velocity increment, changes in those elements scale as $\Delta a /a$ where $\Delta a \sim 10^{-3}$ AU and $a=1.5$ AU. Therefore, escape through violating (\ref{eq:esc_crit}) will occur for ejection speeds two orders of magnitude smaller than those required to leave the \citeauthor{Scholl.et.al2005} stability region. We test this premise further in Section \ref{sec:num_tests}.      
   
The cartesian components of a random vector $\hat{{\rm v}}$ that corresponds to a uniform distribution of points on the unit sphere may be expressed through the parameterisation $\hat{{\rm v}}$=$\hat{{\rm v}}(\theta,\mbox{ }u)$:
 \begin{eqnarray}
 {\rm \hat{v}}_{x}&=&\sqrt{1- u^2}\cos{\theta} \nonumber\\
  {\rm \hat{v}}_{y}&=&\sqrt{1- u^2}\sin{\theta}\\
 {\rm \hat{v}}_{z}&=&u \nonumber
\end{eqnarray}

 where $u$ and $\theta$ are uniformly distributed in the intervals $\left(-1,1\right)$ and $\left(0,2 \pi\right)$ respectively and $u$ is related to the co-latitude $\phi$ through the relationship $u=\cos{\phi}$.
 The conditional probability density function (pdf) of either ${\rm \hat{v}}_{x}$ or ${\rm \hat{v}}_{y}$ given $u$ is given by:
  \begin{equation}
  p({\rm \hat{v}}_{x/y}| u)=\frac{1}{\pi \sqrt{1 - {u}^{2} - {\rm v}^{2}_{x/y}}},\mbox{ }-\sqrt{1 - {u}^{2}} <{\rm v}_{x/y}<\sqrt{1 - {u}^{2}}
  \end{equation}
 
 from which it can be shown that these components of the random vector $\hat{{\rm v}}$ follow the same distribution as $u$.
   
  We now carry out the transformation $(f\mbox{, }u\mbox{, }\theta)$ $\Rightarrow$ $(f\mbox{, }\Delta a^{\ast}\mbox{, }\theta)$
  to obtain the trivariate pdf 
  \begin{equation} \label{eq:pdf_da}
  p(f\mbox{, }\Delta a^{\ast}\mbox{, }\theta)=\frac{|\Delta a^{\ast}|}{8 \pi^2 |K(e, \theta, \phi)| \sqrt{K^{2}(e, \theta, \phi) - {\Delta a^{\ast}}^2}}\mbox{, }|\Delta a^{\ast}| < |K(e, \theta, \phi)|
  \end{equation}
  where $K(e, \theta, \phi) = e \cos{\theta}\sin{f} + (1+e\cos{f})\sin{\theta}$ and $\Delta a^{\ast}$ is the change in semimajor axis that results from a velocity increment $\Delta {\rm v} = |\hat{{\rm v}}| = 1$ .
  
\begin{figure}
\vspace{-3cm}
\centering
\includegraphics[width=120mm,angle=0]{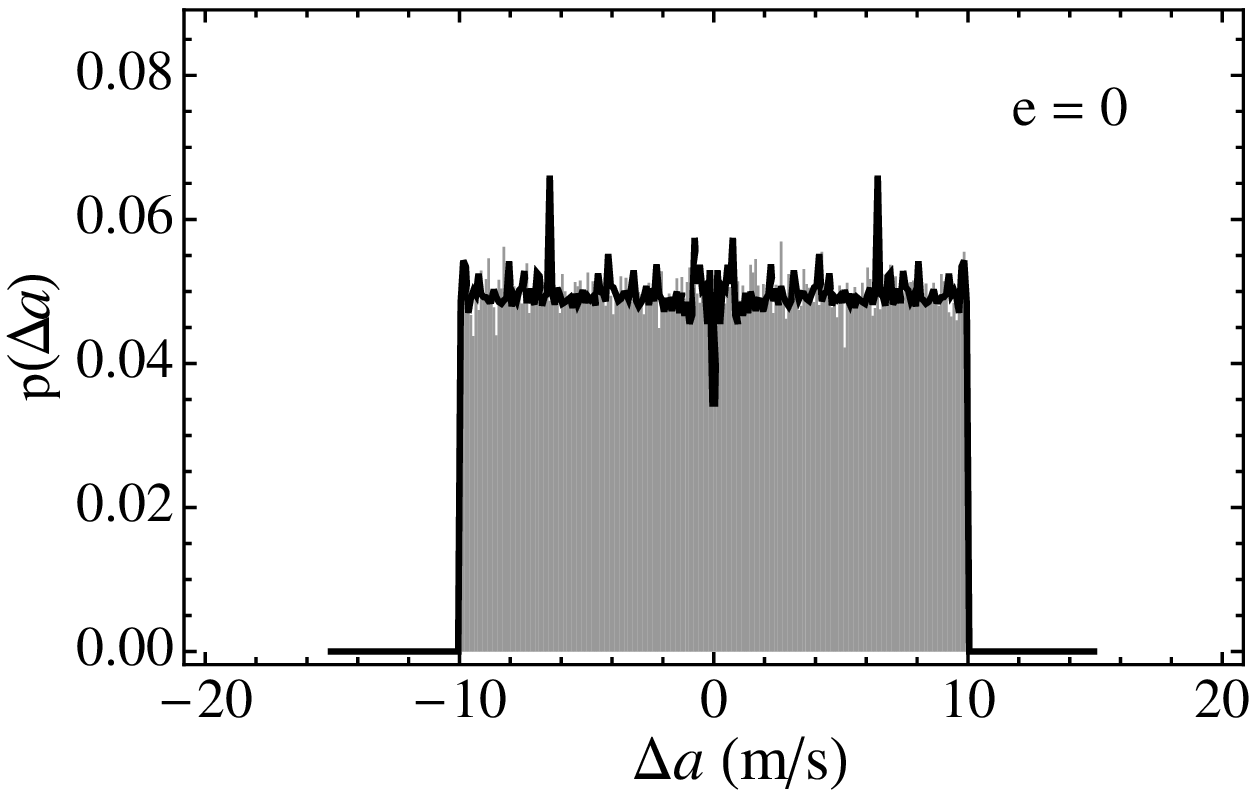}
\includegraphics[width=120mm,angle=0]{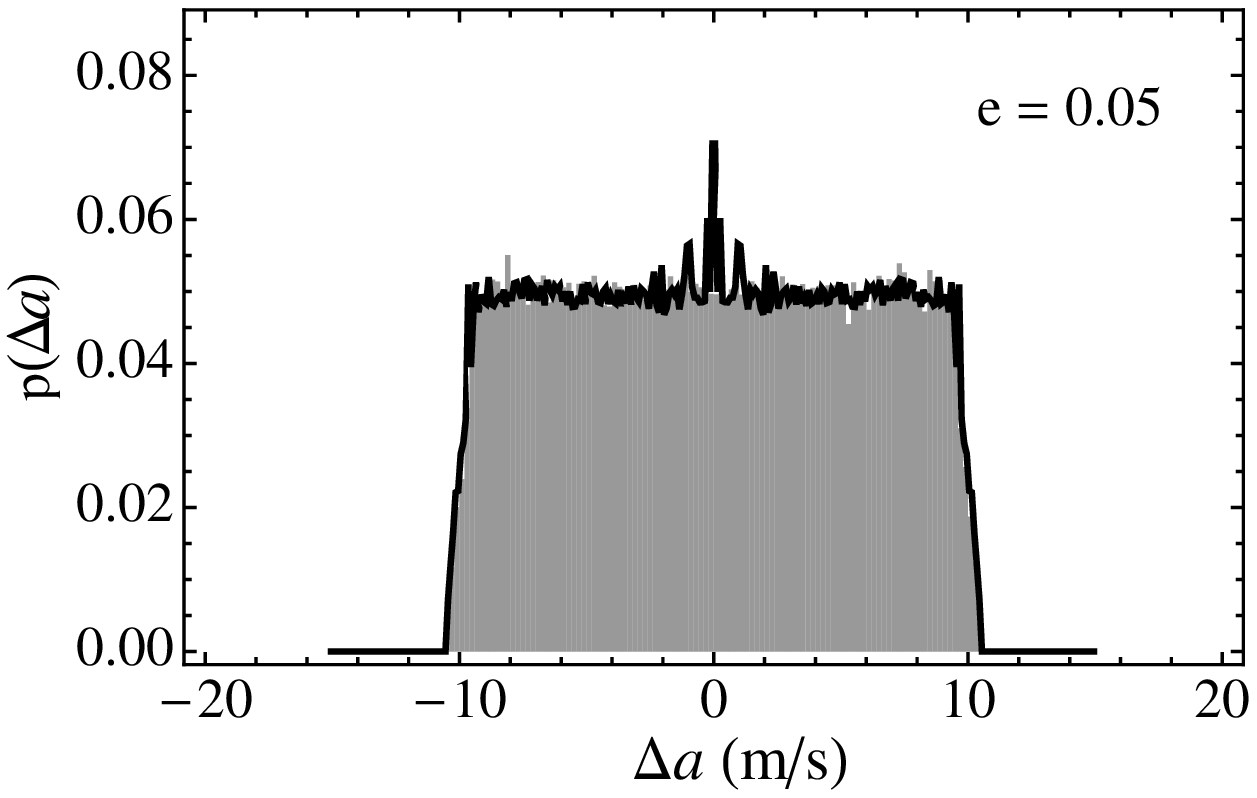}
\includegraphics[width=120mm,angle=0]{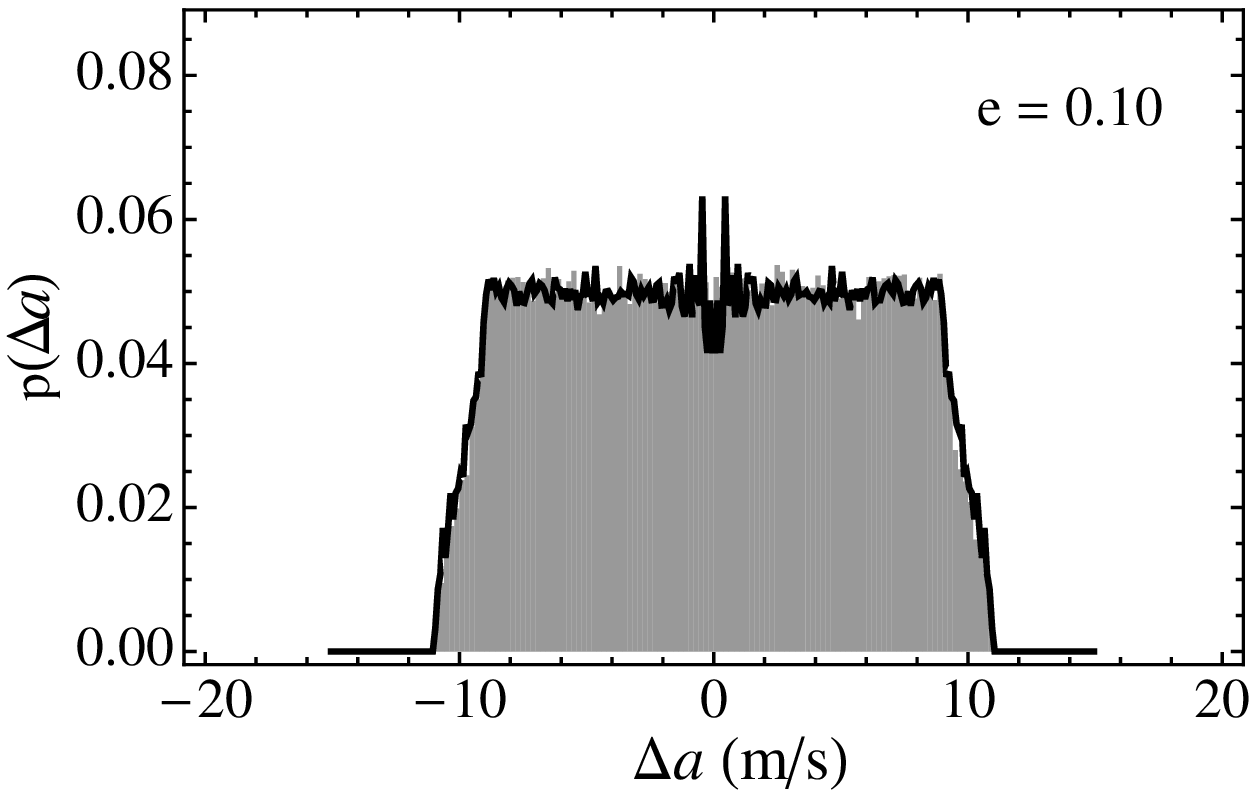}
\caption[Probability density of the change in semimajor axis for a given ejection speed v from Eq.~\ref{eq:pdf_da}. The eccentricity $e$ increases from 0 to 0.1 from top to bottom.]{}
\label{fig:pdf_da}
\end{figure}

  We recover the distribution of $\Delta a$ for a given speed ${\rm v}$ by considering that $\Delta a = \Delta a^{\ast} {\rm v}$ so
  \begin{equation}\label{eq:pdf_da_given_v}
  p(f\mbox{, }\Delta a\mbox{, }\theta \mbox{ }|\mbox{ }{\rm v})=\frac{|\Delta a|}{8 \pi^2 {\rm v} |K(e, \theta, \phi)|  \sqrt{{\rm v}^{2} K^{2}(e, \theta, \phi) - {\Delta a}^2}}\mbox{, }|\Delta a| < {\rm v}\mbox{}|K(e, \theta, \phi)|
 \end{equation}
  
from which one can obtain the pdf of $\Delta a$ for a given speed by double integration. Note that, in deriving (\ref{eq:pdf_da}) and (\ref{eq:pdf_da_given_v}) we have assumed that the true anomaly $f$ varies uniformly in time. Technically this is not correct since the uniformly-varying quantity is the mean anomaly $M$; unless $e=0$, $f$ depends on it in a nonlinear way through Kepler's equation. However, a uniformly-varying $f$ is a good approximation for low-$e$ orbits. We show this in Fig.~\ref{fig:pdf_da} where we plot the distribution of $10^{5}$ random variates of $\Delta a$ (grey bars) for ${\rm v}=10$ m $\mbox{sec}^{-1}$ and for e=0, 0.05 and 0.1 (top, middle \& bottom panels respectively). These were generated from Eq.~\ref{eq:a_dv}, with uniform variates of $\theta$ and variates of $f$ calculated from a Fourier series approximation in $M$, including terms up to $e^{4}$ (e.g.~Murray \& Dermott 1999). The bold line is the double integral of Eq.~\ref{eq:pdf_da} evaluated numerically using the procedure {\tt NIntegrate} in {\it Mathematica}. The  two agree closely, at least up to $e=0.1$, even though $f$ was assumed uniform in producing the bold line.

Fig.~\ref{fig:pdf_da} is also useful in demonstrating how the pdf of $\Delta a$ changes as the eccentricity is gradually increased. For $e=0$, $K=\sin{\theta}$
and $p(\Delta a | {\rm v}) = {\left(2 {\rm v}\right)}^{-1}$ for $| \Delta a| <  {\rm v} $ and $0$ otherwise. This is a good approximation for stable Martian Trojans with $e \lesssim$0.1 \citep{Mikkola.et.al1994}.
 
 If the pdf of ejection speed $p({\rm v})$ is supplied, the pdf of $\Delta a$ is arrived at as the marginal probability density of the joint pdf
 \begin{equation}\label{eq:pdf_da_v}
p(\Delta a\mbox{, }{\rm v}) = p(\Delta a  | {\rm v}) \mbox{ } p({\rm v})
\end{equation} 
by integration.
  
\subsection{Models of fragment ejection speed}
\label{sec:ejection_models}
It is evident from the previous Section, specifically Eq.~\ref{eq:pdf_da_v}, that the distribution of $\Delta a$ for the fragments depends on the adopted distribution for ${\rm v}$. Obtaining physically realistic distributions of ejection speed is an active research area where 
much depends on the - measured or assumed - properties of the impactor and target, quantities poorly constrained even for well-characterised Main Belt asteroids let alone the poorly-observed Martian Trojans \citep{Davis.et.al2002}. To ensure that our conclusions will at least encompass - if not outright represent - the truth, we carry through our computations for two different ejection speed models, those by \citet{Cellino.et.al1999} and \citet{Carruba.et.al2003}. For our purposes, the main difference between the two is that the \citeauthor{Cellino.et.al1999} model relates the ejection speed to fragment size whereas \citeauthor{Carruba.et.al2003} do not.

\subsubsection*{Model A - \citep{Cellino.et.al1999}:}
\citet{Cellino.et.al1999} argued, on grounds of energy equipartition, for a relationship between the ejection speed $V$ and the size $s$ of a fragment of a catastrophic collision of the form
\begin{equation}\label{eq:V}
\log{\left(s/s_{\rm PB}\right)} =2/3 \log V - K^{\prime}
\end{equation}

where $s_{\rm PB}$ is the size of the parent asteroid and $K^{\prime}$ a constant. Comparing with the size-velocity statistics of known asteroid families, they found that this relationship
can adequately represent the {\it maximum} speed (say $V^{\ast}$) for a given fragment size. Furthermore, the additive constant $K^{\prime}$ depends on $r=m_{\rm LR}/m_{\rm PB}$ - the mass ratio of the largest remnant to the parent asteroid - through the empirical law 
\begin{equation}\label{eq:Kprime}
K^{\prime} = 0.21 \log{r} - 1.72
\end{equation}    

On the basis of these findings, they suggested obtaining the ejection speed of a fragment of a given size $s$ as a uniformly random variate between $0$ and the maximum speed from Eq.~\ref{eq:V} once $r$ has been specified. This is our Model A.    

The speed pdf may be obtained from the joint distribution of ${\rm v}$ and $s$
\begin{equation}\label{eq:pdf_v_s}
p({\rm v},s) = p({\rm v} | s) \mbox{ } p(s)
\end{equation} 

where the first term is the speed pdf for a given fragment size while the second is a pdf describing the size distribution of the fragments. The former is assumed to be uniform (cf Eq.~\ref{eq:V}) so
\begin{equation}\label{eq:pdf_v_given_s}
p({\rm v} | s) =  1/ V^{\ast}(s), 0 \leq {\rm v} < V^{\ast}(s)\mbox{.}
\end{equation} 

For a power law size distribution of the form (\ref{eq:size_distr}) with exponent $\alpha$, 
the corresponding pdf can be built by first defining the cumulative distribution function
\begin{equation} \label{eq:cdf_size}
p( < s) =  \left( s_{\rm min}^{-\alpha}  -  s^{-\alpha}\right)  / \left( s_{\rm min}^{-\alpha}  -  s_{\rm LR}^{-\alpha} \right) \mbox{,    } s_{\rm min} \leq s<s_{\rm LR}
\end{equation}
and obtaining the pdf by differentiation.

The joint speed-size pdf as well as the marginal distribution for the speed are given by expressions (\ref{eq:pdf_v_s_cellino}) and (\ref{eq:pdf_v_cellino}) respectively in the Appendix.
For $r \gtrsim 0.8 $ that is applicable to the limiting case $r \simeq 0.9$ discussed at the end of Section~\ref{sec:size}, \citet{Cellino.et.al1999} observed that $K^{\prime}$ no longer follows (\ref{eq:Kprime}) and is approximately constant. The authors
further speculated that this break may be related to the transition between the so-called cratering and catastrophic regimes for the outcome of a collision between asteroids. 
Even though Cellino et al's conclusion was tentative and based on a relatively limited number of families, we do not see a clear distinction between their result and the conventional approach 
of regarding the survival of a 0.5-parent-body-mass fragment as the threshold between the two regimes. It is convenient, therefore, to refer to the two subcases of this model as the cratering and catastrophic regimes respectively. In the cratering subcase, the distributions of interest are then given by (\ref{eq:pdf_v_s_cellino_cratering}) and (\ref{eq:pdf_v_cellino_cratering}). 

The distributions we have just obtained are not robust in the sense that they depend on the - arbitrarily chosen - parameter $s_{\rm min}$. They can be made robust by setting the limiting size to be the minimum {\it observable} fragment size $s_{\rm obs}$ rather than $s_{\rm min}$. In that case, the functional form of the corresponding pdfs is given by (\ref{eq:pdf_v_s_cellino_obs}) and (\ref{eq:pdf_v_cellino_obs}) and similarly for the cratering case. The upper (lower) bounds on the speed (size) domain are now determined by $s_{\rm obs}$ rather than $s_{\rm min}$. 

To arrive at the pdf for $\Delta a$ we multiply the distribution (\ref{eq:pdf_da_given_v}) by either Eq.~\ref{eq:pdf_v_s_cellino_obs} or the equivalent expression for the cratering case and eliminate variables by integration. This results in the branch functions (\ref{eq:pdf_da_cellino_cat}) and (\ref{eq:pdf_da_cellino_cra}) for the catastrophic and cratering cases respectively. {\it Mathematica} procedures to evaluate these expressions  
are available from the authors upon request.

A caveat of the \citeauthor{Cellino.et.al1999} model is related to their interpretation of the orbital dispersion of asteroid family members as solely due to collisional ejection. It was shown subsequently that the gradual action of the Yarkovsky effect modifies significantly this distribution \citep{Bottke.et.al2001} and is an important factor in the constraining of family ages \citep{Vokrouhlicky.et.al2006, Spoto.et.al2015}. For this reason, Model A should overestimate ejection speeds and the semimajor axis dispersion of the fragments. The implications for our conclusions regarding the possibility that the Eureka family formed by a collision are discussed at the end of this Section. 

\subsubsection*{Model B - \citep{Carruba.et.al2003}:}
The alternative method by \citet{Carruba.et.al2003} (Model B) generates synthetic asteroid families from collisions at the catastrophic disruption limit
where the critical value $Q^{\ast}$ of the specific energy is obtained as a function of the bulk density $\rho_{\rm PB}$ and radius $R_{\rm PB}$ of the parent body respectively:
\begin{equation}\label{eq:qstar_carruba}
Q^{\ast}=0.4*\rho_{\rm PB}*R^{1.36}_{\rm PB} \mbox{.}
\end{equation}
The separation velocity ${\rm v}_{\infty}$ of the fragments is obtained through
\begin{equation}\label{eq:v2inf_carruba}
{\rm v}^2_{\infty}={\rm v}^2_{\rm ej} - {\rm v}^{2}_{\rm esc}
\end{equation}

where 
\begin{equation}\label{eq:v2esc_carruba}
{\rm v}^{2}_{\rm esc}=1.64*G*M_{\rm PB}/R_{\rm PB}\mbox{.}
\end{equation}
The ejection velocities follow a maxwellian pdf with a quadratic mean of 
\begin{equation}\label{eq:v2ej_carruba}
<{\rm v}^{2}_{\rm ej}> = Q^{\ast}*f_{\rm KE}\mbox{,}
\end{equation}
where $f_{\rm KE}$ represents the fraction of the impact energy that goes into the kinetic energy of the fragments.

The distribution of the separation speed ${\rm v}_{\infty}$ for escaped fragments may be defined through the ejection speed pdf $p_{ej}(.)$ as
\begin{equation}\label{eq:escape_distr}
p_{\infty}=p_{ej}({\rm v}^{2}_{\infty}+{\rm v}^{2}_{esc})/\sqrt{1+{\rm v}^{2}_{esc}/{\rm v}^{2}_{\infty}}\mbox{.}
\end{equation}

This is an un-normalised distribution function mapping all values ${\rm v}_{ej}>{\rm v}_{esc}$  to a positive ${\rm v}_{\infty}$.
Consequently, it integrates not to unity but, rather, the fraction of ejection speeds in $p_{ej}$ that result in escapes. This integral is given by
$\Gamma \left(1/2,\kappa^{2}\right)/\sqrt{\pi}+2\sqrt{\kappa/\pi} \exp^{-\kappa^{2}}$ where $\kappa^{2}={\rm v}^{2}_{esc}/2 \sigma^{2}$, $\sigma^{2}=3 <{\rm v}^{2}_{\rm ej}>$ and $\Gamma$ is the incomplete Gamma function. It approaches unity if $<$${\rm v}^{2}_{\rm ej}$$> \gg {\rm v}^{2}_{esc}$, in other words when all of the fragments escape.
From Eq.~\ref{eq:qstar_carruba}, \ref{eq:v2esc_carruba} \& \ref{eq:v2ej_carruba} and for typical parameter values - $R_{\rm PB}=1$ km, $\rho_{\rm PB} =2500$ kg $\mbox{m}^{-3}$, $f_{\rm KE}=0.1$ and $Q^{\ast}\sim 10^3$ J $\mbox{kg}^{-1}$ - 
we get  $<$${\rm v}^{2}_{\rm ej}$$> / {\rm v}^{2}_{esc} \simeq 100$ therefore $p_{\infty}(.)$ is effectively a maxwellian pdf of the same form as $p_{ej}(.)$.
 
As this speed is assumed independent of the size $s$, the corresponding pdf for the change in semimajor axis is obtained by multiplying the pdf (\ref{eq:pdf_da_given_v}) (or its simplified form for small $e$) with $p_{ej}(.)$ and integrating from $|\Delta a|$ to infinity to obtain:
 
 \begin{equation}\label{eq:da_distr_carruba}
p_{CA04}(\Delta a)= 1/ \sqrt{2 \pi \sigma^{2}} e^{- {\Delta a}^{2}/2 \sigma^{2}}
\end{equation}

 ie a Gaussian. 

\subsection{Numerical checks}
\label{sec:num_tests}
Before proceeding further, we wish to check that our assumed mechanism of escape - dispersion of the semimajor axis $a$ of the fragments and escape if $|a - a_{\rm Mars} | > a_{\rm crit}$ - is correct. This is done in two stages:
Firstly, we  generated a set of fragment speed-size pairs by sampling the distributions (\ref{eq:pdf_v_given_s}), (\ref{eq:cdf_size}) and (\ref{eq:escape_distr}). These pairs were used to calculate a set of heliocentric state vectors for the fragments where the speeds, applied along uniformly random directions, were vectorially added to the ephemeris velocity of Eureka at 0000 UTC on 1st July 2000 retrieved from JPL HORIZONS \citep{Giorgini.et.al1996}. The position vectors of the fragments were assumed identical to that of Eureka. These state vectors were then numerically integrated for 2000 yr with a 2nd order Bulirsch-Stoer scheme available within the {\sl MERCURY} package \citep{Chambers1999} and with a model that included the 8 major planets. We find that fragments that escape from the Trojan region do so typically within the first few tens of years from the beginning of the integration. 

\begin{figure}
\vspace{-3cm}
\centering
\includegraphics[width=75mm,angle=0]{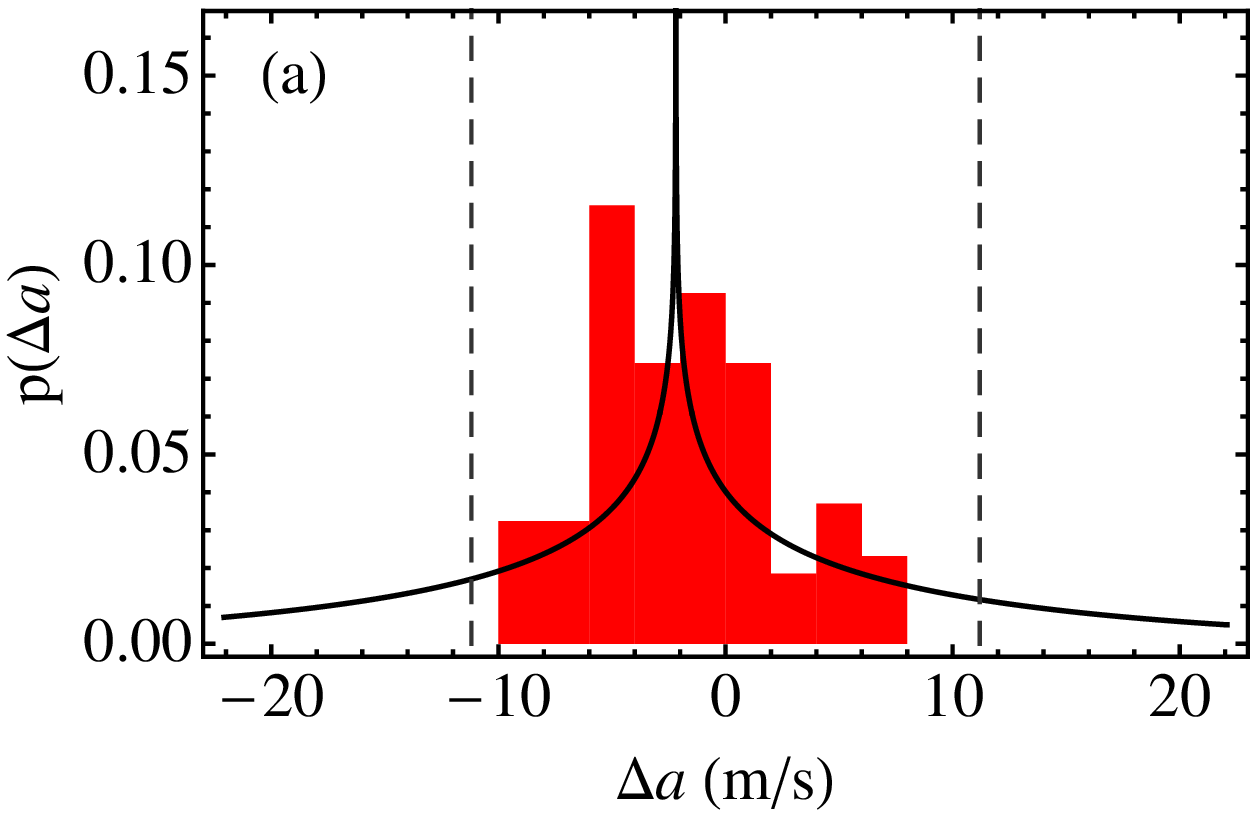}\mbox{ }\includegraphics[width=75mm,angle=0]{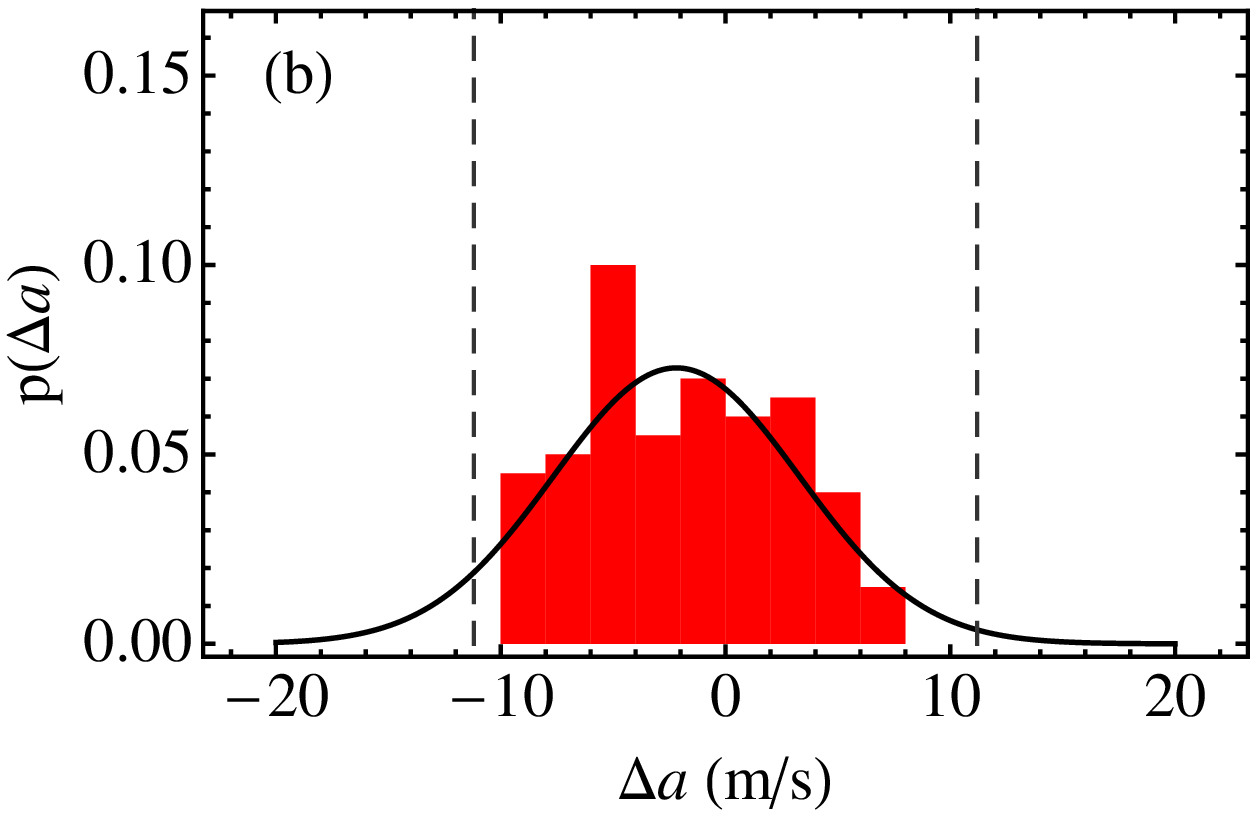}
\caption[Retention of trojan fragments as a function of the change in semimajor axis $\Delta a$ from the numerical tests described in Section~\ref{sec:num_tests} and for the ejection laws of (a) \citet{Cellino.et.al1999} in the cratering regime, and (b) \citet{Carruba.et.al2003} with $f_{\rm KE}=0.01$, $Q^{\ast}=10^{3}$ J $\mbox{kg}^{-1}$. The corresponding pdfs are plotted as bold curves.The dashed vertical lines indicates the critical value of $\Delta a$ for a Trojan to escape. Values of parameters common in both cases were $r=0.8$, $\alpha=2.5$ and $s_{\rm obs}=70$m.]{}
\label{fig:pdf_sims_da_celino_carruba}
\end{figure}

In the first test, we calculate the quantity $\Delta a = a-a_{\rm Mars}$ averaged over the first 20 years of the integration for each of the fragments that persist as Trojans. The averaging is done to eliminate short-period planetary perturbations in the variation of $a$ and isolate the change due to the coorbital potential. The distribution of the values of  $<$$\Delta a$$>$ is shown in Fig.~\ref{fig:pdf_sims_da_celino_carruba} where we have superposed the respective analytical form from the previous Section. In the left panel, the cratering case for Model A with $s_{\rm obs}=70$m and $s_{LR}=1300$m was used. In the right panel, Model B with $f_{\rm KE}=0.01$ and $Q^{\ast}=10^{3}$ J $\mbox{kg}^{-1}$ was used. In both cases we assumed a size distribution of the form (\ref{eq:cdf_size}) with $\alpha=2.5$. The distributions are offset from centre since ${< \Delta a >}_{Eureka} \simeq -2$ m $\mbox{s}^{-1}$\footnote[$\star$]{$\Delta a$ may be represented in units of speed by omitting the multiplicative factor in Eq.~\ref{eq:a_dv}}. The approximate number of surviving Trojan fragments in either case is $\sim$100. The vertical dashed lines represent the relationship $|< \Delta a >| = a_{\rm crit}$. We observe that, although the theoretical speed distribution of fragments extends further than this boundary, the distribution of persisting Trojan fragments in either case does not. It appears therefore that condition (\ref{eq:esc_crit}) may be used with confidence as a criterion for Trojan retention (or escape). It is also noteworthy that, although the two pdfs are of different shape, the corresponding sample distributions are not too dissimilar. This is partly due to the moderate sample size but also in that both pdfs are supposed to describe the same physical phenomenon: the ejection of asteroidal fragments from a collision near the limit of catastrophic disruption of the target. 

In the second test, we measure how well our analytical model predicts the fraction of escaping (or, equivalently, remaining) fragments. We carry out numerical simulations as before, but with two different sets of values of the parameters that control the velocity distributions of the fragments. At the same time, we calculate the expected fraction of escaping Trojans by integrating the form of $p({\Delta a})$ appropriate for each case across the domain defined by (\ref{eq:esc_crit}). These parameter sets correspond to different simulation runs and the results are shown in Table~\ref{tab:tests}. Again, the statistics from the numerical runs agree well with the analytical model predictions. 

\subsection{ Use of the formal variance}

The principal feature of the family that we wish to reproduce is its compactness, defined here as the degree of dispersion in the libration amplitudes or, equivalently, the semimajor axis of its members.
An appropriate statistic for this purpose is the variance (or its square root, the standard deviation), equal to the order 2 moment of a zero-mean distribution i.e. $E[{\Delta a}^{2}]$.  
For Model A, multiplying (\ref{eq:pdf_da_cellino_cat}) with ${\Delta a}^{2}$ and integrating yields (\ref{eq:var_da_cellino_cat}) and a similar expression (\ref{eq:var_da_cellino_cra}) for the cratering case. For model B, the variance is simply the square of the scale parameter $\sigma$. 

To obtain the distribution of $\Delta a$ for the sample, we carry out numerical integrations of the asteroids' motion using the mixed variable symplectic (MVS) option in MERCURY with an integration time step of 4d, an output step of 128d and a duration of $\sim 6 \times 10^{3}$ yr. This timescale is equivalent to $\sim 5$ libration cycles of the guiding centre \citep{Mikkola.et.al1994} and, at the same time, considerably shorter than the timescale for secular evolution of the orbit ($\sim 5 \times 10^{4}$ yr). From the integration output, we form the complex quantity
$$
\left(1+a - a_{\rm Mars} \right)\exp{ {\rm i}\left(\lambda - \lambda_{\rm Mars}\right)}\mbox{,}
$$
 then use the Frequency Modified Fourier Transform \cite[FMFT;][]{SidlichovskyNesvorny1997} to calculate the libration amplitudes of $a - a_{\rm Mars}$ and $\lambda - \lambda_{\rm Mars}$ as the semi-axes of an ellipse. This is a valid approximation to the guiding centre path around the equilibrium point for small libration amplitude \citep{MurrayDermott1999}, this being the case here. The location of the libration centre is not fixed at $60^{\circ}$ but remains a free parameter to be estimated from the fit. This is done to account for excursions of the libration centre from $\pm60^{\circ}$, expected for non-circular, non-planar orbits \citep{NamouniMurray2000}. In Table \ref{tab:proper} we compare our estimates of $\Delta \lambda$ with those from \citet{Cuk.et.al2015}. Differences in the $\Delta \lambda$ estimates for the individual asteroids range from under a degree (Eureka, 2007 $\mbox{NS}_{2}$) to $\sim 5^{\circ}$ (2011 $\mbox{SC}_{191}$). The sample distribution moments are, however, very similar; the respective means and standard deviations are $2.95\pm0.58$ m $\mbox{s}^{-1}$ and $2.79\pm0.50$ m $\mbox{s}^{-1}$. Therefore, this procedure may also be thought of as alternative method to calculate Trojan libration amplitudes \citep{Milani1993,Cuk.et.al2015} and we can use the new estimates to test the correlation found by \citeauthor{Cuk.et.al2015} with the mean orbital inclinations. We find, as those authors did, that the two are negatively correlated but more weakly so ($R^{2}$=$0.17$ vs $0.83$ for that first set of libration amplitudes). Therefore the \citeauthor{Cuk.et.al2015} conclusion stands, but should be re-examined as new discoveries add to the inventory of known Trojans. 

\begin{figure}
\vspace{-3cm}
\centering
\includegraphics[width=75mm,angle=0]{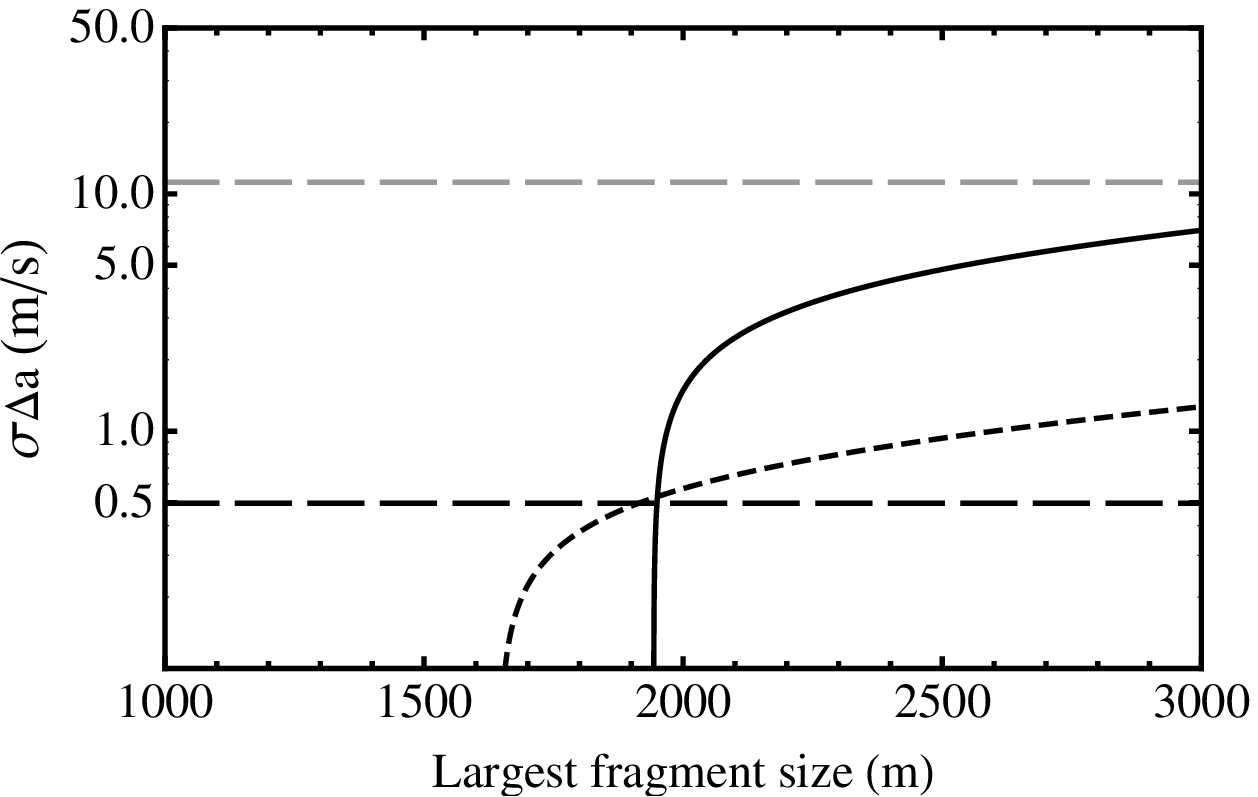}\mbox{ }\includegraphics[width=75mm,angle=0]{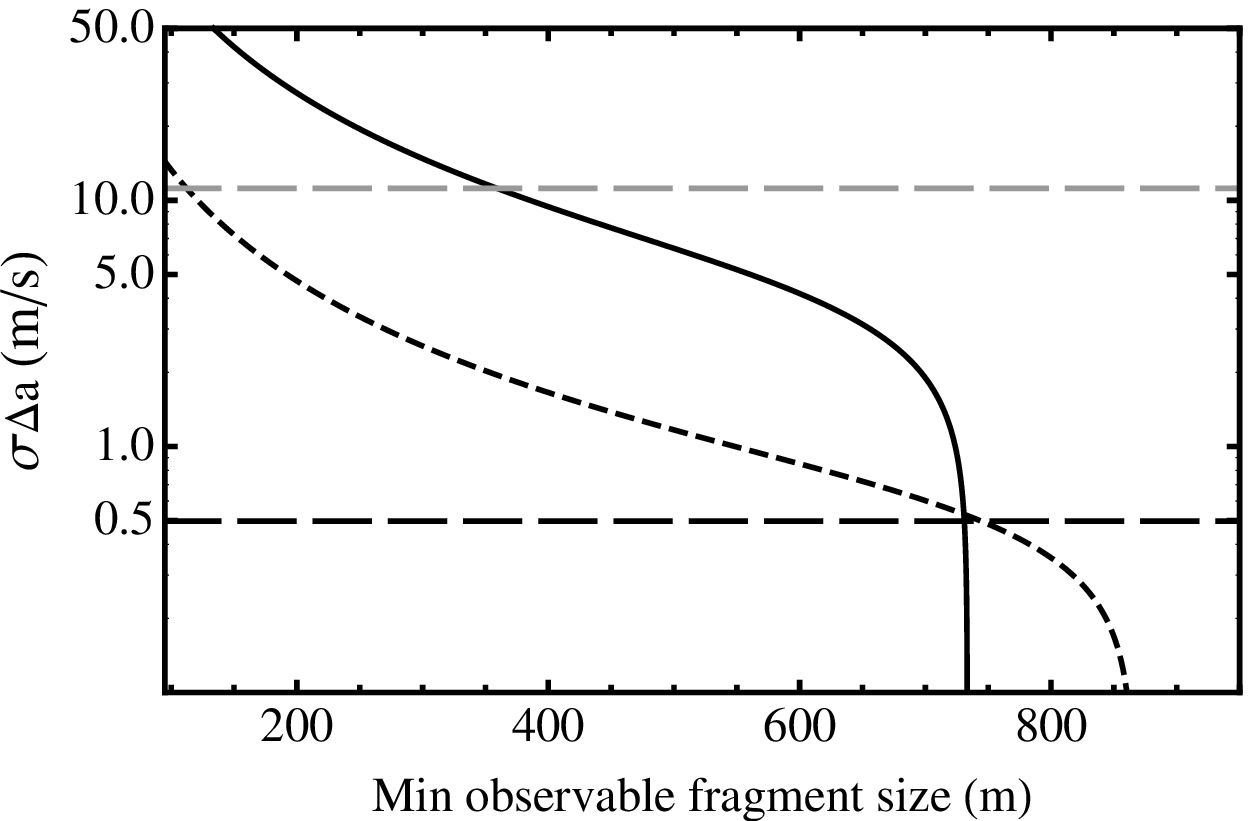}
\caption[Square root of the variance as a function of the largest remnant size (left) and the min observable fragment size (right) for the ejection laws of \citet{Cellino.et.al1999} in the cratering (bold line) and catastrophic regimes (short dashed line) for $r=0.5$,  $s_{\rm LR}=3000$m and $s_{\rm obs}=s_{\rm LR}/30$, and the model by \citet{Carruba.et.al2003} with $f_{\rm KE}=0.1$, $Q^{\ast}=10^{4}$ J $\mbox{kg}^{-1}$. The corresponding pdfs are plotted as bold curves.The long dashed horizontal line indicates the critical value of $\Delta a$ for a Trojan to escape.]{}
\label{fig:variance_sdi_sobs_celino_carruba}
\end{figure}

It is useful, at this point, to quantify the sensitivity of the variance to the different model parameters, particularly for Model A where the variance depends on them in a nontrivial algebraic fashion. By experimenting with different values, we find that our model is most sensitive to the size of the minimum observable fragment $s_{\rm obs}$ and the size of the largest fragment $s_{\rm LR}$. In Figure~\ref{fig:variance_sdi_sobs_celino_carruba} we show $\sqrt{V[{\Delta a}]}$ for the ejection laws of \citet{Cellino.et.al1999} in the cratering (bold line) and catastrophic regimes (short dashed line) as functions of these two parameters where we have set $r=0.8$. The two dashed horizontal lines indicate the critical value of $\Delta a$ for a Trojan to escape (grey line)  and the root-mean-square of the sample for the Eureka family (black line). Further, we have set $s_{\rm obs}=750$m in the left plot and  $s_{\rm LR}=1900$m in the right. The \citeauthor{Carruba.et.al2003} model (Model B) is independent of fragment size and generally predicts a somewhat higher dispersion than observed, for instance $\sigma_{\Delta a}=1.73$ m $\mbox{s}^{-1}$ for $f_{\rm KE}=0.01$ and $Q^{\ast}=10^{2}$ J $\mbox{kg}^{-1}$. $\sigma_{\Delta a}\simeq0.5$ m $\mbox{s}^{-1}$ would require reducing either $f_{\rm KE}$ or $Q^{\ast}$ by an order of magnitude to values not easy to reconcile with present understanding of the physics of asteroid collisions \cite[e.g.][]{Jutzi.et.al2009}. Therefore we focus our attention on the size-dependent Model A from this point onwards.  

\begin{figure}
\vspace{-3cm}
\centering
\includegraphics[width=150mm,angle=0]{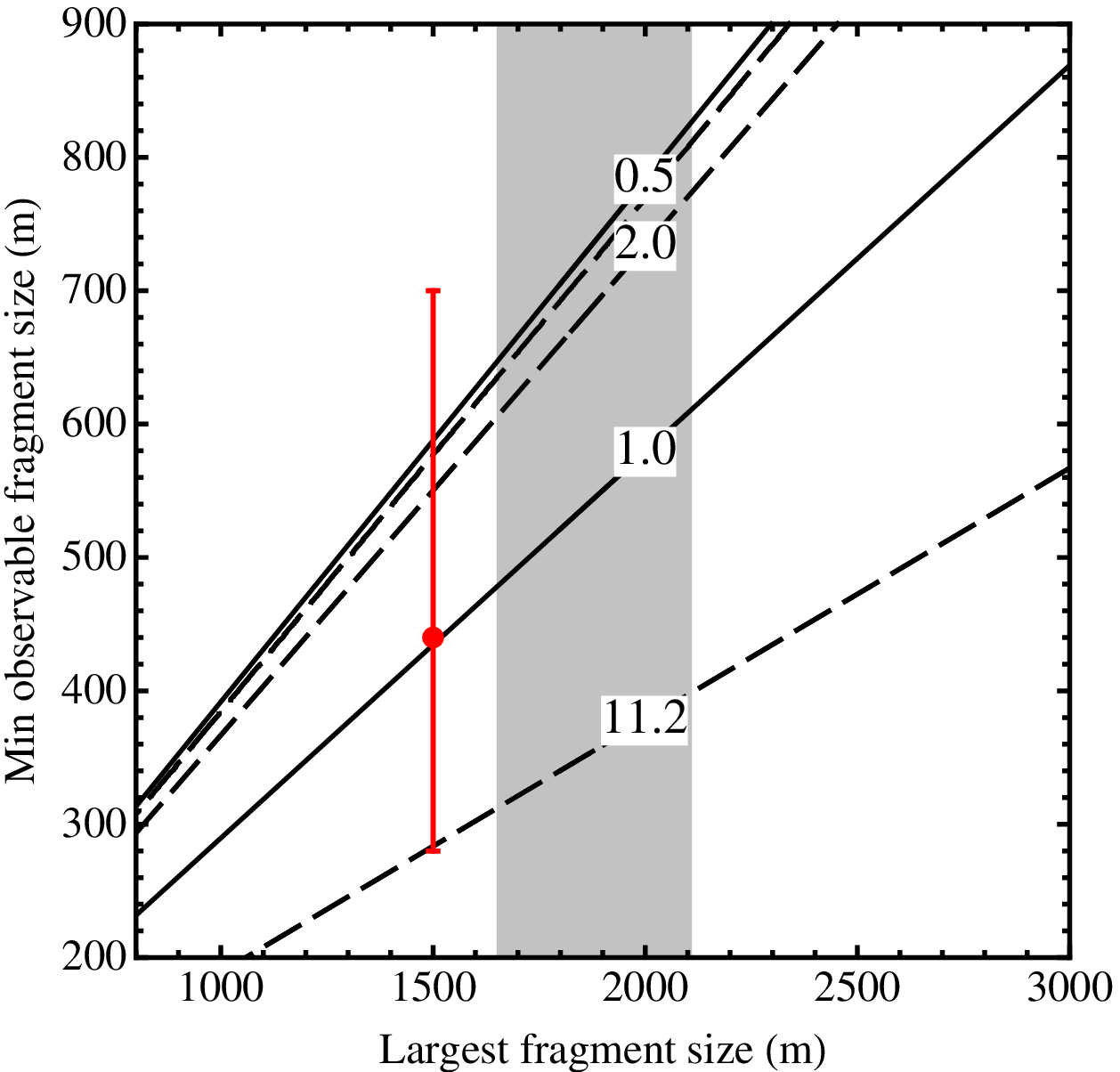}
\caption[Contours of constant $\sigma$ as a function of the largest remnant size and the minimum observable fragment size. See main text for details.]{}
\label{fig:variance_slr_sobs_cellino}
\end{figure}

As one might expect, the law for the catastrophic case generally predicts a higher dispersion for given values of $s_{\rm obs}$ and $s_{\rm LR}$ than the law for the cratering case. Apparently, retainment of Trojan fragments occurs over a wide range of values for both parameters. The same cannot, however, be said for families of fragments as compact as the Eureka family. In this instance, values of $\sigma_{\Delta a}$ up to 2 $\times \sigma_{\rm Eureka}$ are achieved for $s_{\rm LR}=1600-2000$m in the left panel and  $s_{\rm obs}=600-850$m in the right. Interestingly, as $\sigma_{\Delta a}$ decreases the two profiles approach each other. This is not unexpected since $r=0.8$ signifies the break between the cratering and catastrophic impact regimes in the \citeauthor{Cellino.et.al1999} model. This close agreement is valid over the entire range of values of $s_{\rm obs}$ and $s_{\rm LR}$ relevant to this problem. Fig.~\ref{fig:variance_slr_sobs_cellino} shows contours of constant variance as a function of these two parameters. The vertical gray band represents the nominal WISE constraint on Eureka's size. The $0.5$ m $\mbox{s}^{-1}$ contour, co-incident for the two regimes, corresponds to the size range $650\mbox{m}<s_{\rm obs}<840\mbox{m}$ or $18.4<H<17.9$. 
It is noteworthy that the cratering regime (continuous line) is more sensitive to an increasing $\sigma_{\Delta a}$ than the catastrophic one (dashed line). Whereas quadrupling the standard deviation to $2$ m $\mbox{s}^{-1}$ does not significantly affect the latter profile, a change of half that size shifts the cratering profile by ${\Delta s}_{\rm obs}=-170$m. 
In other words, allowing the target value of $\sigma_{\Delta a}$ to vary results in different ranges in ${s}_{\rm obs}$ between the cratering and catastrophic regimes. It is then useful, as a consistency check, to ask the question of what is the minimum observable size for a Martian Trojan at present? Although we are not aware of a specific study addressing the level of completeness of Martian Trojan surveys, this subject has been investigated for the near-Earth Object population. The distribution of NEO orbits is somewhat different than that for Martian Trojans as NEOs can be observed closer to the Earth than MTs, favouring the detection of smaller asteroids. On the other hand, a NEO would spend most of the time well outside the orbit of Mars where it may be unobservable.

\citet{HarrisDAbramo2015} calculated NEO survey completeness as a function of the absolute magnitude $H$ over the last 20 yr. The red point and error bar in Fig.~\ref{fig:variance_slr_sobs_cellino} represent the median (50\%) and interquartile range (25\% $\rightarrow$ 75\%) of the distribution of their ``C20'' quantity from their Table 2. Our analysis of the family dispersion implies somewhat higher values for ${s}_{\rm obs}$ but there is overlap with the 0.5 m $\mbox{s}^{-1}$ contour for 650-700m ($H \sim 18.3$) objects. Invoking our earlier argument, it appears easier to accommodate a cratering regime scenario, i.e. one where the ejection velocity is independent of $s_{\rm LR}$. An actual Trojan limiting magnitude significantly fainter than that for NEOs would be more favourable towards a catastrophic regime scenario and vice versa for a brighter $H$. 
To distinguish between these two scenaria would require a dedicated study of the present completeness level of MTs and/or dedicated searches for Martian Trojans to drive 50\% completeness to the few hundred m level. We see, for instance, that for $s_{\rm obs}=300$m ($H\simeq 20$) the variance for the catastrophic regime becomes 11.2 m $\mbox{s}^{-1}$, several times higher than for the cratering regime and equal to the entire width of the Trojan region.

Finally, if, as discussed in Section 4.2, Model A overestimates fragment ejection speeds, the contours in Fig.~\ref{fig:variance_slr_sobs_cellino} would be shifted down and to the right. This would mean that the family is complete to a smaller size and that the parent body was larger. In fact, the collisional hypothesis may be dismissed altogether if this overestimate is sufficiently severe to move the contours far from the gray bar. However, the fact that our two ejection speed models yield similar results in the marginally catastrophic regime (Fig.~\ref{fig:pdf_sims_da_celino_carruba} and see discussion above) suggests that this is not the case.  

\section{Calculation of collisional lifetime} 
\label{sec:probability}

Computation of the collisional lifetime requires (i) a statistical model of the orbits and sizes of the impactors, and (ii) a physical description of the conditions for catastrophic disruption of the target. The impactor populations relevant to Martian Trojans are: Mars-Crossing Main Belt asteroids (hereafter referred to as Mars Crossers - MCs), Near Earth Objects (NEOs), Jupiter Family Comets (JFCs) and other Martian Trojans (MTs).

Describing the size distribution for each class of impactors by the power law (\ref{eq:size_distr}), estimates of the exponent $\alpha$ for MBAs in the literature vary over a wide range \cite[1.3-3.0;][]{Neukum.et.al1975,Belton.et.al1992,Ivezic.et.al2001,Tedesco.et.al2005,Bottke.et.al2007} while for NEOs, $\alpha \approx 2$ 
\citep{Bottke.et.al2002,StuartBinzel2004}. For reference, the expected value for a collisionally relaxed population is 2.5\footnote{Formally true for a size-independent critical energy density $Q^{\ast}$ \cite[see][and references therein]{Davis.et.al2002}} \citep{Dohnanyi1969}. 

\begin{figure}
\vspace{-9cm}
\centering
\includegraphics[width=150mm,angle=0]{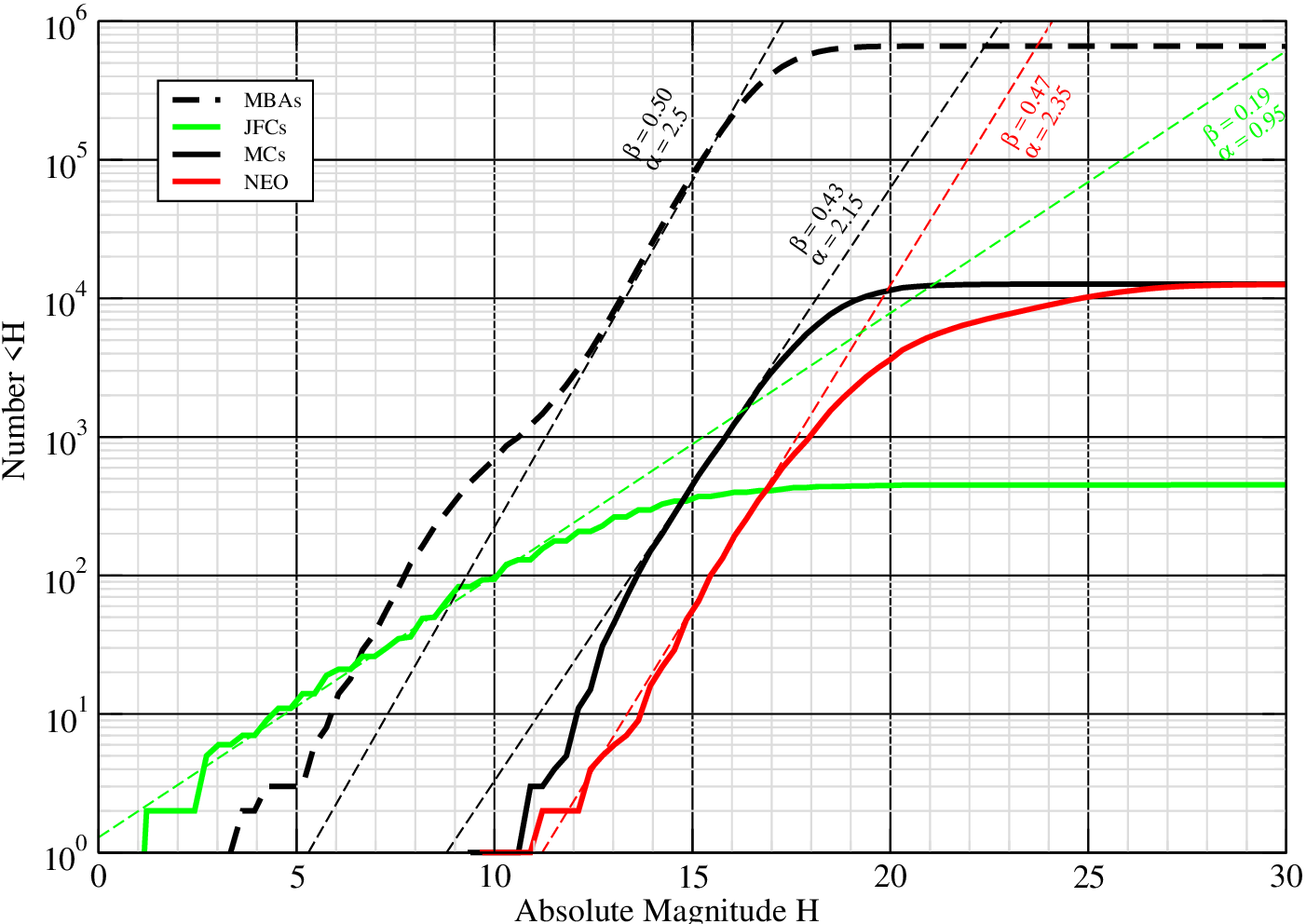}
\caption[Cumulative number of objects with absolute magnitude less than H versus H, for MBAs (dashed black line), MCs (solid black line), NEAs (solid red line) and JFCs (solid green line). Data was obtained from the Minor Planet Center on 27/05/15. The straight dashed lines represent power law fits over a suitable range of sizes.]{}
\label{fig:N_vs_h}
\end{figure}

Fig.~\ref{fig:N_vs_h} shows the cumulative number of bodies with absolute magnitude less than $H$ for MBAs (black dashed line), MCs (black solid line), NEAs (red line) and JFCs (green line) (data retrieved from the Minor Planet Center on May 27, 2015). The rollover for fainter objects is due to the loss of detection efficiency for those asteroids \cite[eg][]{DeMeoCarry2013}. Since the majority of impactors that can lead to catastrophic disruption of Martian Trojans are considerably smaller than those represented in the observational data, we extrapolate from suitable power laws to account for their contribution in our model. The straight lines in Fig.~\ref{fig:N_vs_h} represent power-law fits to the MPC data over specific magnitude intervals. Note that the contribution from JFCs is at least an order of magnitude lower than those of the other two source populations for objects with H$>$22. For this reason we neglect the contribution of JFCs in our model. In addition, the slope for MCs is somewhat shallower than that for MBAs and closer to that of NEOs. It is, therefore, reasonable to utilise the same power law dependence for the MC and NEO populations and express the former as a multiple of the latter i.e.~$N_{MC}(>D) = 8 N_{NEO}(>D)$. For $N_{NEO}(>D)$ we adopt a functional form similar to that proposed by \citet{StuartBinzel2004}: $N (>D) = 1090 \mbox{ }D[\mbox{km}]^{-\alpha}$ for NEOs with $\alpha$ a free parameter in our model.

The distribution of potential impactor orbits is used to determine the mean intrinsic probability of collision $P$ (in units of $\mbox{km}^{-2}$ $\mbox{yr}^{-1}$ $\mbox{impactor}^{-1}$, the average number per year of close approaches between the centres of the target and an impactor within a distance of $1$ km) and the distribution $\psi (U)$ of the impact speed. The latter may be expressed as $\psi(U)$=$dn/dU$, where $dn$ is the mean number per year of close approaches within a distance of 1 km with relative velocity between $U$ and $U+dU$. The distribution of the selected orbits depends on $H_{cutoff}$, the absolute magnitude cutoff we apply. This parameter is unrelated to the power law fits to the size distributions discussed above. 

\begin{figure}
\vspace{-3cm}
\centering
\includegraphics[width=150mm,angle=0]{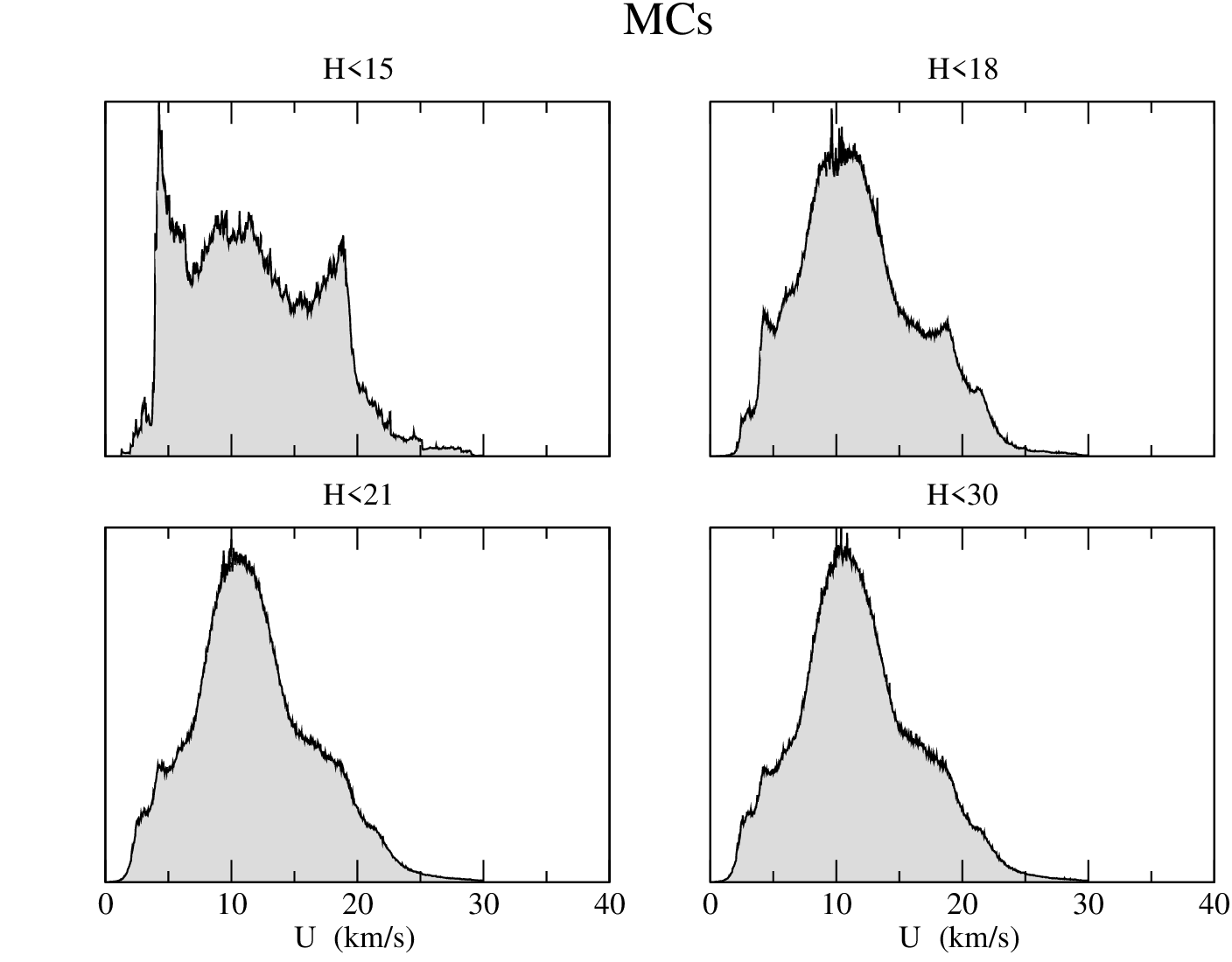}
\caption[Impact speed distributions corresponding to the Mars Crosser impactor population (MC)  as a function of the imposed size cutoff. The vertical and horizontal scales are the same for all panels in the Figure.]{}
\label{fig:V_vs_h_mc}
\end{figure}

To understand how choosing a particular cutoff value may affect the impact velocity distribution, we have calculated $\psi(U)$ for $H_{cutoff}=15$, $18$, $21$ and $30$.
For the MC population (Fig \ref{fig:V_vs_h_mc}) we find that, considering only the brightest objects (H$<$15) yields a trimodal $\psi(U)$ with peaks at 4.5, 10 and 19 km $\mbox{sec}^{-1}$. Including orbits for fainter objects results in a more unimodal distribution for $\psi (U)$ peaking at $10$ km $\mbox{sec}^{-1}$ with most ($>90\%$) of the power between 5 and 20 km $\mbox{sec}^{-1}$. It is worth noting that this is somewhat higher than the case for the Main Asteroid Belt \cite[e.g.][]{FarinellaDavis1992}. For NEOs, there are essentially no differences between the obtained impact velocity distributions except for a gradual reduction of the dispersion about the mode as progressively fainter objects are included. Fewer differences still are apparent in the case of the JFC population.

\begin{figure}
\vspace{-3cm}
\centering
\includegraphics[width=150mm,angle=0]{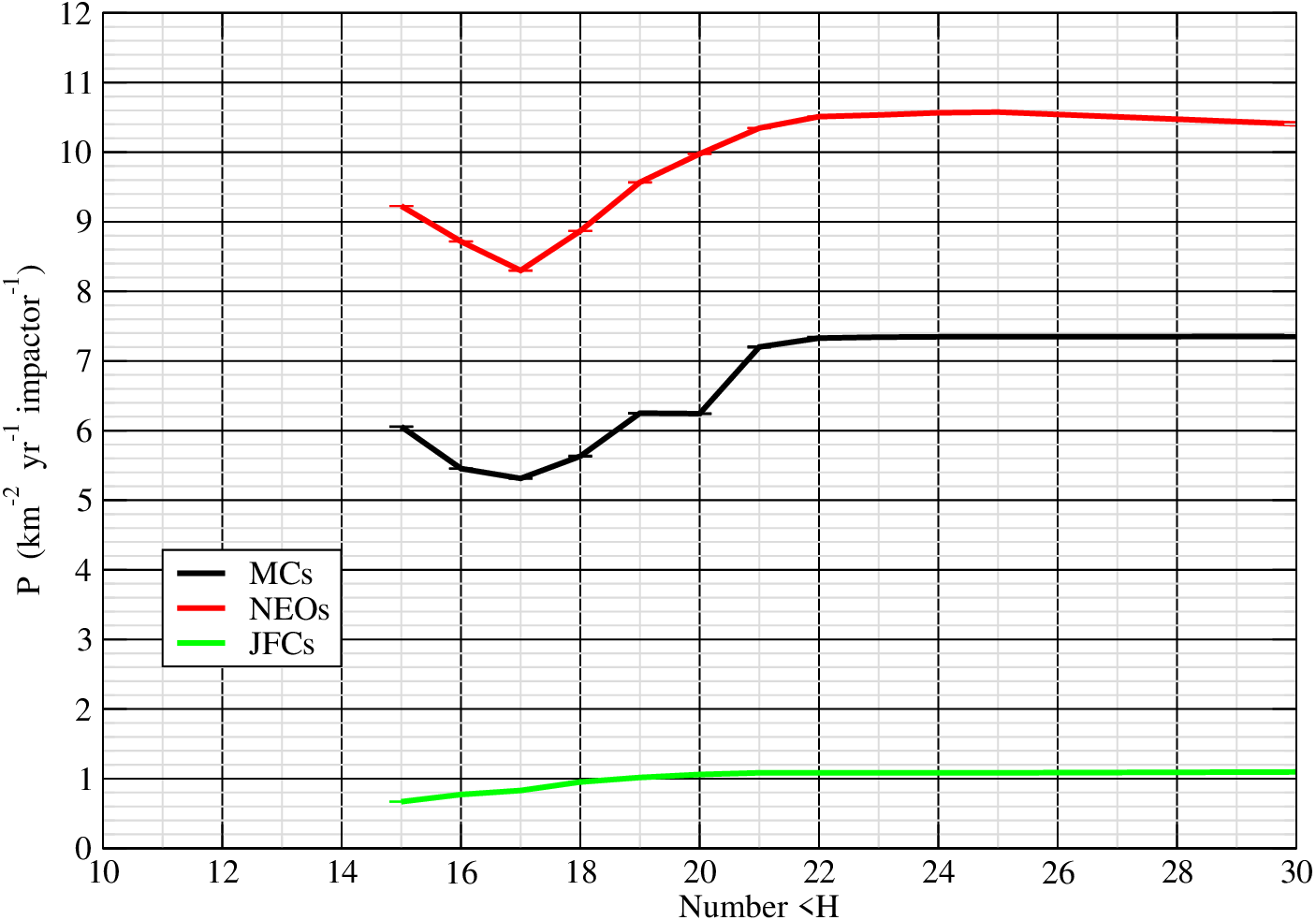}
\caption[Impact probability for the different populations of impactors as a function of  $H_{cutoff}$.]{}
\label{fig:Pimp_vs_h}
\end{figure}

Fig.~\ref{fig:Pimp_vs_h} shows that the dependence of $P$ on $H_{\rm cutoff}$ is similar for the MC and NEO populations with a shallow minimum at $\sim$17 and a flattening out for objects fainter than 22. 
Different choices for the value of $H_{cutoff}$ entail variations of the computed intrinsic collision probability of 20\%-30\%. A better estimation of $P$ would require the modelling of the debiased orbital distribution 
of the different impactor populations, i.e. corrected for observational selection effects, a task beyond the scope of the present work. The uncertainty in $P$ is a relatively minor factor affecting the final estimate of the collisional lifetime compared to the impactor size distribution and the energy required to disrupt the target (see below).
In light of these facts, we have adopted $H_{cutoff}$=18 since choosing a higher value does not drastically change the result.
The value of $P$ for JFCs is about an order of magnitude smaller than for MCs and NEOs and probably contains many fewer impactors with respect to those two populations (Fig.~\ref{fig:N_vs_h}). This further justifies our decision to neglect the contribution of JFCs in our model. 

Inter-Trojan collisions, which we have not considered so far in our computations, may increase the value of $P$ over and above what is shown in Fig.~\ref{fig:Pimp_vs_h}. This increase should, however, be fairly small according to the following argument: For the Trojans listed in \citet{Christou2013} the mean intrinsic probability of collision is $\sim$ $68 \times 10^{-18}$ $\mbox{km}^{-2}$ $\mbox{yr}^{-1}$ and the mean impact velocity is 11 km $\mbox{s}^{-1}$. The value of $P$ is an order of magnitude higher than that for collisions with MCs while for objects of the Mars Trojans' sizes, at 11 km $\mbox{s}^{-1}$, the critical projectile size for catastrophic disruption should be $\sim 20$m, corresponding to $H=26-27$. In order for the contribution from Trojan-Trojan collisions to be comparable to the contribution from MCs, the number of Trojans with $H < 26-27$ should be at least 10\% of MCs or a few times $10^{7}$. 
However, a power law extrapolation assuming $\beta=0.5$, and that the $\mbox{L}_{5}$ population is complete up to $H=17$ where $N(H) = 2$, shows that there should be only $2 \times 10^{5}$ Trojans down to that size. Therefore we believe that small Trojans, if they exist, make a negligible contribution to the impactor population.

The parameter determining the destruction of the target is the so-called specific impact energy
\begin{equation}  \label{eq:Q}
Q = 1 / 2 (m/M) U^{2}
\end{equation}

where $m$ and $M$ are respectively the masses of the impactor and the
target and $U$ is the impact speed.

\begin{figure}
\vspace{-3cm}
\centering
\includegraphics[width=150mm,angle=0]{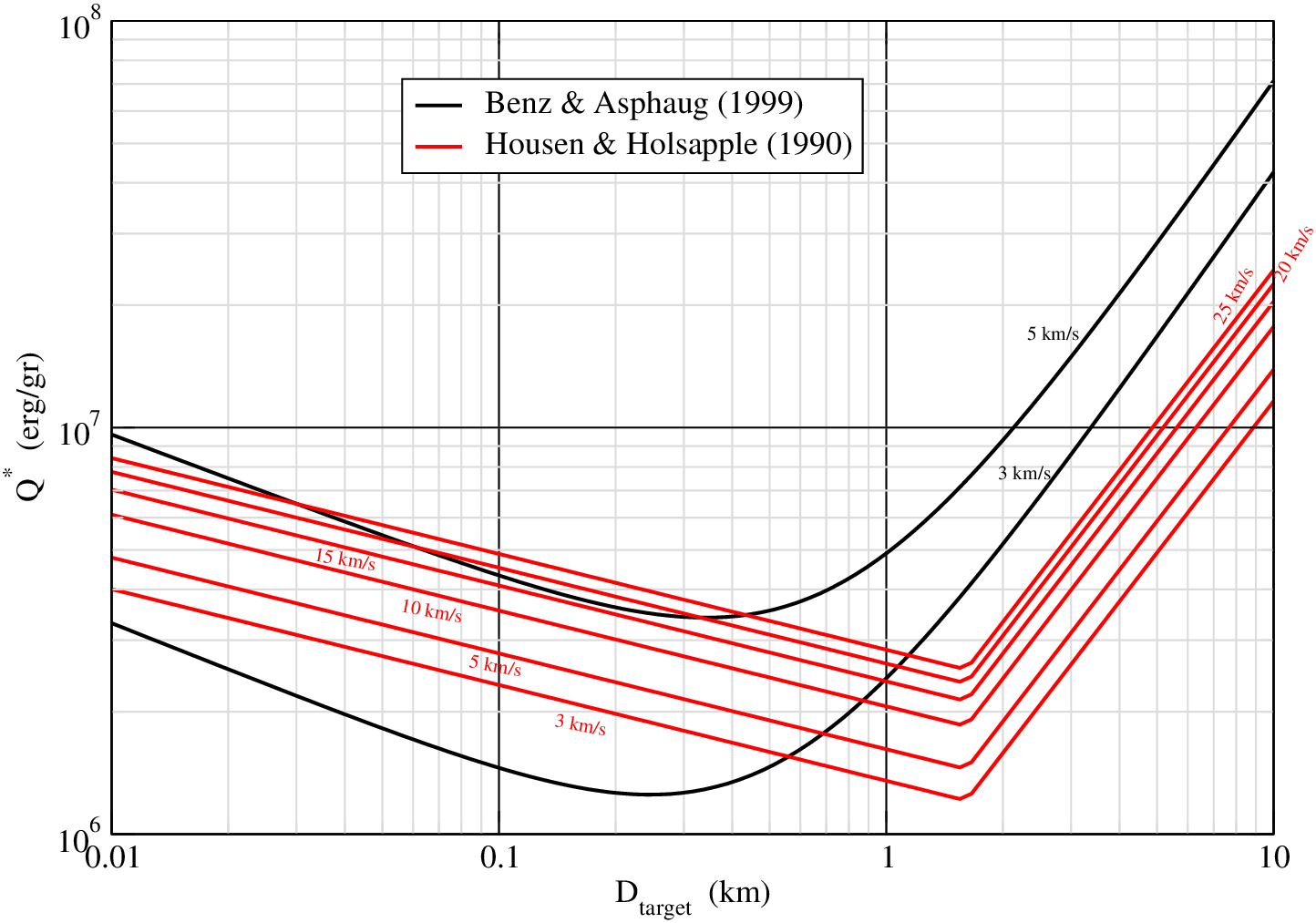}
\caption[Dependence of the critical specific energy $Q^{\ast}$ on target diameter $D$ according to \citet{BenzAsphaug1999} and 
\citet{HousenHolsapple1990}.]{}
\label{fig:Qstar_vs_D}
\end{figure}

An impact is typically termed ``catastrophic'' if it results in the fragmentation and dispersion of the target with the largest fragment containing at most 50\% of the parent body's mass. 
This occurs when $Q > Q^{\ast}$ where the critical value $Q^{\ast}$ depends - apart from the material properties of the impactor \& target - on the size $D$ of the target and the impact velocity $U$. If such an event was responsible for the Eureka family, the size of the parent body would still be only 25\% larger than Eureka, about 2 km. We considered the following two models for $Q^{\ast}$ (Fig.~\ref{fig:Qstar_vs_D}): the one by \citet[][hereafter BA99]{BenzAsphaug1999}, developed specifically for collisions among MBAs of basaltic composition and for impact speeds of 3 and 5 km $\mbox{s}^{-1}$, and the other by \citet[][hereafter HH90]{HousenHolsapple1990}, for which $Q^{\ast}$ can be computed for any value of the target diameter and impact velocity.
Even though fragmentation models are based on rigorous physical assumptions, the outcomes they predict are still characterised by large uncertainties.  
As a case in point, we note that BA99 does not reproduce exactly the results of HH90 for $U$ = 3 and 5 km $\mbox{s}^{-1}$.

All calculations are carried out for the following reference orbit for the Martian Trojan targets: $a$ $=$ 1.523840 au, $e$ $=$ 0.05588 and $I$ $=$ 21.30 deg. 
The numerical computation of the intrinsic probabilities of collision and distributions of the impact velocity has been done using the method of \citet{DelloroPaolicchi1998}.
We further assume that the population of the potential impactors as well as that of the targets are in a steady state, that the size and orbit 
distributions of potential impactors are uncorrelated and that their angular elements (longitudes of the node and pericentre) are uniformly distributed.
This last assumption follows the \"Opik/Wetherill formalism where the semimajor axes, eccentricities and inclinations are held fixed and the pericenter arguments 
and node longitudes vary uniformly. Under this particular condition the general formalism of \citet{DelloroPaolicchi1998} is equivalent to that of \citet{Bottke.et.al1994}, and reproduces 
the results of the latter work.

\begin{figure}
\vspace{-3cm}
\centering
\includegraphics[width=150mm,angle=0]{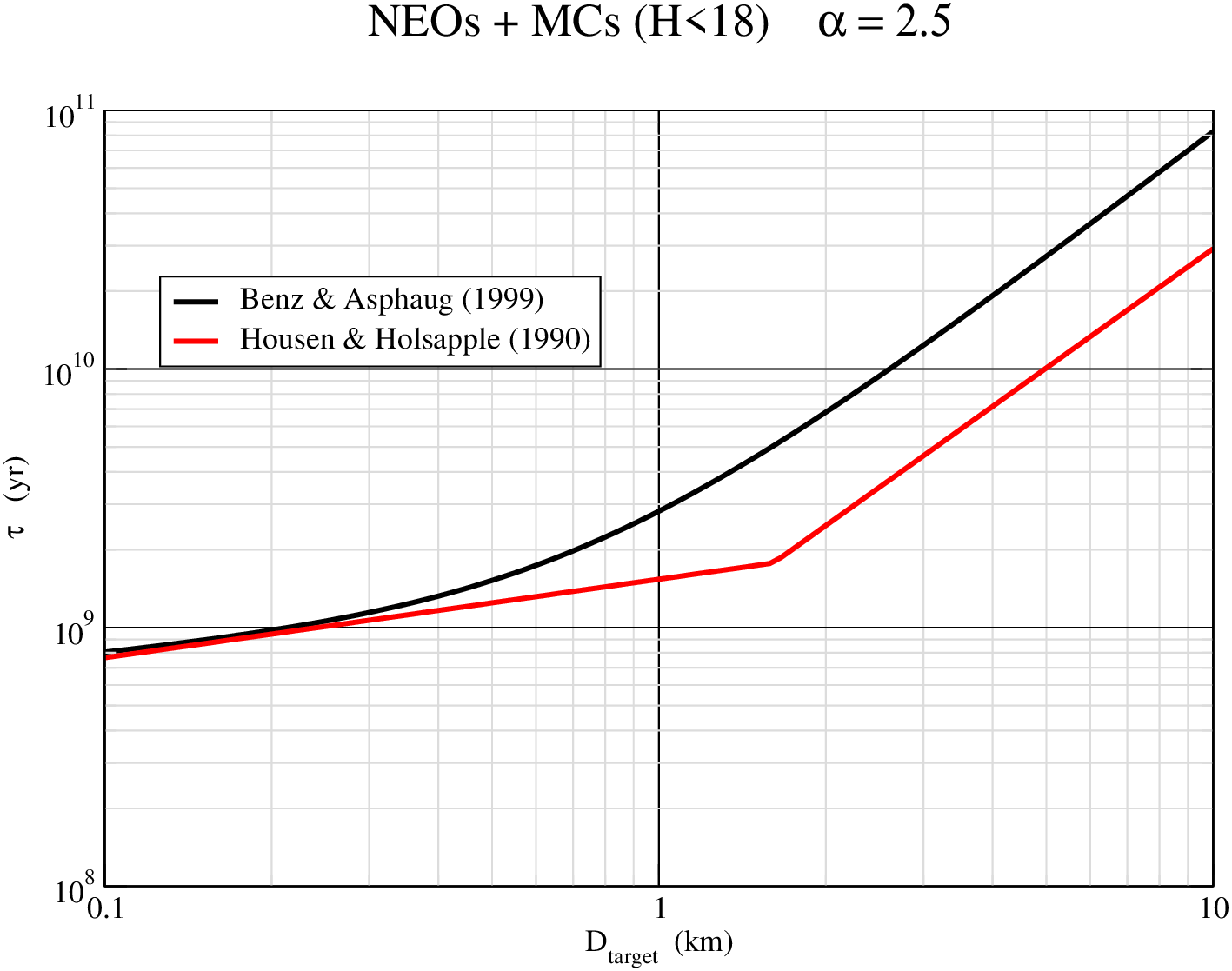}
\caption[Collisional lifetime of Martian Trojans as a function of size for two different collisional fragmentation models and for $\alpha=2.5$.]{}
\label{fig:tau_vs_d}
\end{figure}

\begin{figure}
\vspace{-3cm}
\centering
\includegraphics[width=150mm,angle=0]{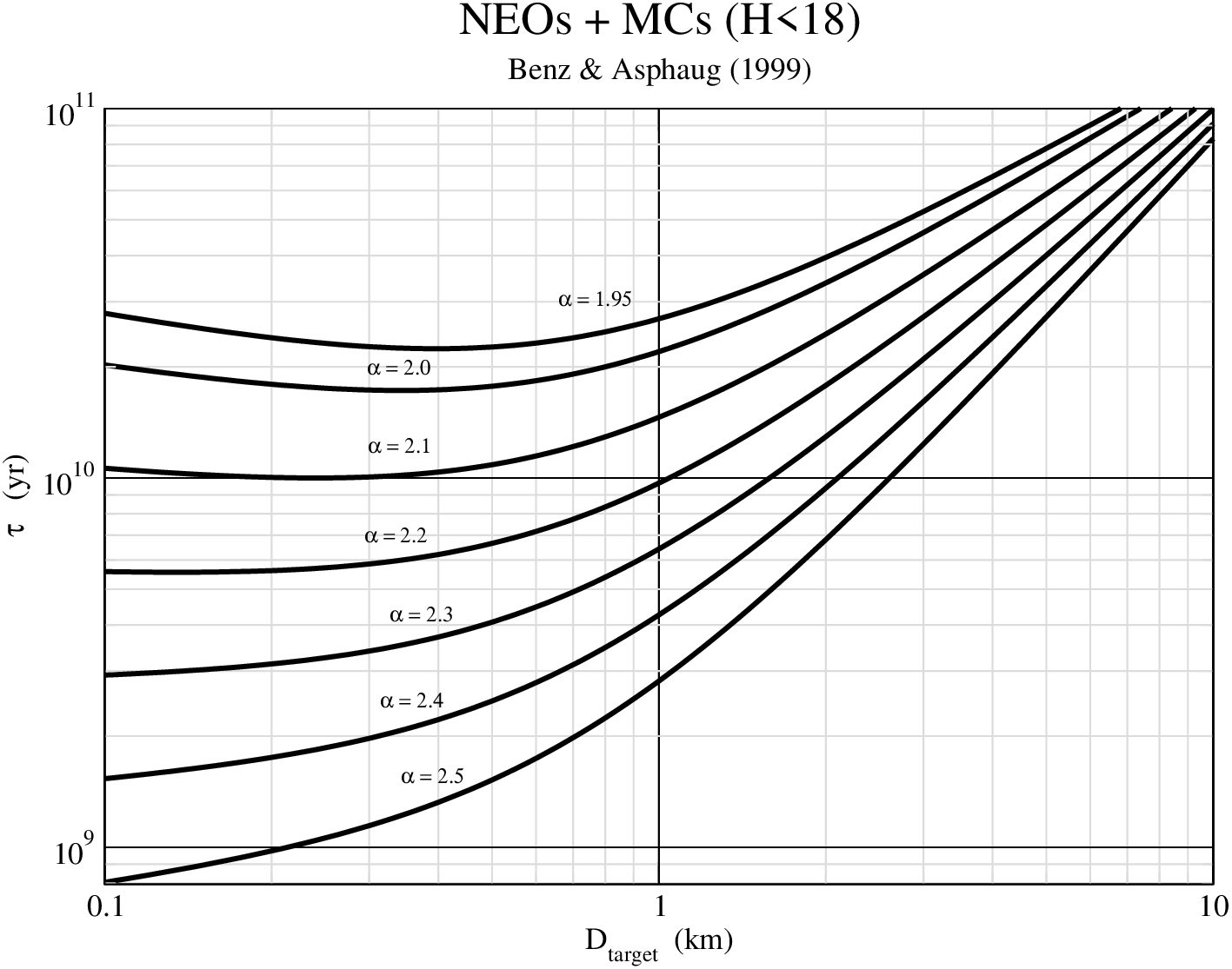}
\caption[Collisional lifetime of Martian Trojans as a function of size for different size distribution index $\alpha$ for the impactor population and assuming the fragmentation model of \citet{BenzAsphaug1999}.]{}
\label{fig:tau_vs_d_alpha}
\end{figure}

Fig.~\ref{fig:tau_vs_d} shows the collisional lifetime of a Martian Trojan as a function of size for a specific value of the impactor size distribution index ($\alpha=2.5$) and for the collisional fragmentation models of HH90 (red curve) and BA99 (black curve). We observe that the two models agree up to $D\simeq250$ m where $\tau \simeq 10^{9}$ yr. They then diverge until they become again parallel to each other for $D\gtrsim2$ km so that the estimated $\tau$ for \citeauthor{HousenHolsapple1990} is about a third of that for \citeauthor{BenzAsphaug1999} above that size. This is due to the somewhat higher critical energy density $Q^{\ast}$ in the latter model (Fig.~\ref{fig:Qstar_vs_D}). The dependence of $\tau$ on the impactor size distribution index is illustrated in Fig.~\ref{fig:tau_vs_d_alpha} where we show curves for different values of $\alpha$ between
2.5 and 1.95 in increments of $0.1$ for the \citeauthor{BenzAsphaug1999} model. A shallowing of the impactor size distribution increases $\tau$ particularly for the smaller sizes with the effect that a 1 km object survives for 4.5 Gyr if $\alpha \lesssim 2.4$ and the same holds at the small end of the size range if $\alpha \lesssim 2.25$. These lifetimes are significantly longer than for similar-sized objects in the Main Belt where a 0.1 and 1 km object will disrupt every $140$ and $450$ Myr respectively \citep{Farinella.et.al1998}.
The probability that a Martian Trojan will suffer at least one catastrophic impact during a time interval $\Delta t=\tau$ is $1-e^{-1} \sim 63\%$ while a 95\% chance of at least one impact requires $\Delta t \simeq 3 \tau$. Therefore, for $\alpha$ significantly less than 2.5 and under the \citeauthor{BenzAsphaug1999} model, a 2 km Martian Trojan such as the Eureka parent body, 1998 $\mbox{VF}_{31}$ and 1999 $\mbox{UJ}_{7}$ may not have suffered a single catastrophic collision over the age of the solar system. A first-order independent check of this result is to consider the 3 largest Martian Trojans as different outcomes of the same Poisson process. The likelihood function of the observation of a cluster around Eureka (a ``hit'') but no clusters around the Trojans 101429 and 121514 (two ``misses'') is
\begin{equation}  \label{eq:likelihood}
L (\tau) = e^{-2 \Delta t / \tau} \left(1 -  e^{- \Delta t / \tau} \right) 
\end{equation}
which has a maximum at $\tau^{\ast}= \Delta t / \ln \left(3/2\right)$. For $\Delta t = 4.5$ Gyr, $\tau_{\ast}$ evaluates to $11$ Gyr, in good agreement with the model prediction in Fig.~\ref{fig:tau_vs_d_alpha}.
Under the alternative model of \citeauthor{HousenHolsapple1990}, $\tau$ is revised downwards by a factor of about 3. In that case, the survival time for a 2 km object is  2-3 Gyr, about half the age of the solar system and still quite long compared to their main belt counterparts. Incidentally, this confirms the conjecture of \citet{Rivkin.et.al2003}.

The values of $\tau$ we have computed represent mean time intervals between catastrophic collisions. There remains the possibility that the family formed in a cratering event, where $r>0.5$ and $Q<Q^{\ast}$. In that case, it can be shown that the mean interval $\tau_{c}$ between such events relative to the same interval $\tau$ for marginally catastrophic disruptions is

\begin{equation}
\tau_{c} / \tau \simeq {\left(Q_{c} / Q^{\ast} \right)}^{\alpha/3}\mbox{.}
\end{equation}
 
 Assuming $r \simeq 0.9$ for the Eureka family, the simulations of \citet{BenzAsphaug1999} suggest $Q_{c} / Q^{\ast} = 0.1-0.2$ \cite[see also][]{LeinhardtStewart2011}
so for $\alpha$ between 2 and 2.5, we obtain $\tau_{c}=\left(0.15-0.30\right) \tau$ or $\geq$1 Gyr for a 2 km asteroid.
Therefore, and contrary to our findings for catastrophic events, there is a high likelihood ($\sim$99\%) that Eureka has suffered one or more such impacts over the age of the solar system. 
 
 \section{Synthesis}
 The analysis done in Section \ref{sec:dispersion} shows that fragment confinement in the Martian Trojan clouds following a collisional event is possible. However, the compactness of the Eureka family within said cloud places severe constraints on such an origin scenario; we showed that only a marginally catastrophic or sub-catastrophic (i.e.~cratering) event can reproduce the observed libration amplitude dispersion among family members. The collisional scenario becomes altogether untenable if, for the reasons discussed in Section~\ref{sec:ejection_models}, the size-dependent ejection speed law by \citet{Cellino.et.al1999} used in our orbital dispersion model significantly overestimates fragment ejection speeds. Indeed, the subsequent orbital evolution of the fragments under gravitational perturbations and a Yarkovsky-like drag force, would lead to a dispersion in libration amplitude of a compact group of Trojans after about a Gyr \citep{Cuk.et.al2015} so either the observed dispersion is (i) solely due to the ejection velocity field from a collision, in which case the family must be younger than a Gyr, or (ii) a combination of collisional dispersion and orbital evolution, suggesting ejection speeds lower than inferred from the observed dispersion of the orbits. It appears, therefore, that the only viable scenario from the fragment dispersion modelling is a cratering event sometime in the past 1 Gyr where the fragments were ejected at or near the parent body's escape speed. This is consistent with the findings of \citeauthor{Cuk.et.al2015} and also supported by our estimate of the mean cratering rate for the Eureka parent body (Section~\ref{sec:probability}). However, such low-energy impacts tend to produce second largest fragments smaller by an order of magnitude or more in size than the largest remnant \citep{Durda.et.al2007,Benavidez.et.al2012}. In the case at hand, (311999) 2007 $\mbox{NS}_{\rm 2}$ is $\sim$1/3 the size of Eureka (Table~\ref{tab:eureka family}), apparently too large to be a product of a cratering event. Possible workarounds to this apparent quandary include an oblique impact \citep{Durda.et.al2007} and that the parent body is a strengthless rubble pile \citep{Benavidez.et.al2012}, both of which act to reduce the disparity between fragment sizes.
The shallow size distribution of the family, taken at face value, would also argue against a collisionally-generated population. This argument is, in principle, strengthened by the observation that models of large-$r$ size distributions that consider the finite dimensions of the parent body and the collisional fragments tend to follow steeper size distributions than those with low $r$ \citep{Tanga.et.al1999}. 
One should keep in mind, however, that the slope of the size distribution is currently underestimated due to as-yet-undiscovered Trojans at the faint ($H$$\sim$20) end of the observed population (Section~\ref{sec:size}). As evidence of significant observational incompleteness for the fainter Trojans, we point to two recent discoveries near $\mbox{L}_{5}$, 2016 $\mbox{AA}_{165}$ ($H$$=$20.4) and 2016 $\mbox{CP}_{31}$ ($H$$=$19.7). Although not yet confirmed as Martian Trojans or members of the Eureka family, their orbital semi-major axis, orbital location about $60^{\circ}$ behind Mars and orbital eccentricity and inclination make them strong family member candidates, hopefully to be confirmed when recovered at a future apparition. 

 \section{Conclusions and Discussion}
The main findings of this paper may be summarised as follows:
 
\begin{enumerate} 
\item
The collisional lifetime of the present population of Martian Trojans is of order of Gyr, longer than that of same-sized Main Belt asteroids. 
\item 
The orbital distribution of Eureka family members is consistent with family formation in the recent $\sim20\%$ of the solar system's history from a sub-catastrophic collision. This result is at odds with the sizes of the largest family members. 
\item  
The family size distribution is shallower than that expected for a group of collisional fragments and similar to that of small MBAs. Its present usefulness as a discriminator between competing formation models is, however, limited by the small available sample size. 
\end{enumerate}

We can conceive a number of distinct formation mechanisms for this family, none of them in complete agreement with all the data at our disposal.
They are:
\renewcommand{\labelitemi}{$-$}
\begin{itemize}
\item \emph{Origin from a single collision.} This is the simplest possible model. It would be in agreement with the rate of sub-catastrophic or cratering impacts on Mars Trojans, and not in dramatic disagreement with the apparently shallow size distribution, because we have already seen that the latter could be a result of inventory incompleteness. This scenario would also be exciting given the taxonomy of Eureka family members, implying an olivine-rich composition. Such asteroids are
rare in the main belt, suggesting that most of these objects have been ``battered to bits'' \citep{Burbine.et.al1996,Chapman1997} long ago. If this is true, finding a family of olivine-rich objects in a region where the collisional evolution has been very mild
would suggest an origin from an old parent body, possibly a survivor from an early generation of 
planetesimals.
There are two problems encountered by this model. One is the large sizes of the family asteroids (311999) 2007 $\mbox{NS}_{2}$ and (311999) 2007 $\mbox{DH}_{47}$ relative to Eureka, a feature not generally seen in numerical simulations of sub-catastrophic impacts. Rather than a fatal flaw in the collisional hypothesis, this could be indicative of the specific impact conditions or the internal structure of the parent body. The other is the strong compactness of the Eureka family, requiring original ejection velocities from the parent body very close to the escape velocity. This is somewhat problematic in a collisional environment
in which, according to our results, the collision rate is rather low but the collision velocities tend to be high, producing 
complete fragmentation of the target body by a smaller projectile.
\item \emph{Origin from rotational fission}. YORP-induced rotational fission is apparently in agreement with 
available dynamical and physical evidence. However, even this explanation of the origin of the Eureka family is not exempt from problems. In particular, the small collision rate in the Mars Trojan region would suggest that these asteroids could experience nearly pure YORP-driven evolution, only very mildly disturbed by collisions modifying the shape and pole orientation. If this is true, the possibility that during its history Eureka began undergoing fission events very recently (to justify the compactness of the family) appears unlikely. The present family could in principle be the outcome of the last episode of a series of fission events experienced by Eureka since its origin. However, no trace of these more dispersed ``YORPlet'' asteroids from previous episodes are observable today, suggesting at least some upper limit to the frequency of YORP-driven fissions, taking into account the timescale of Yarkovsky-driven dispersion computed by \citet{Cuk.et.al2015}.  Incidentally, this is also a problem for the cratering impact formation scenario if they occur every few $10^{8}$ yr or so.

Here, a clue may be the shorter ($\lesssim 10^{9}$ yr) collisional lifetime of Trojans a few 100s of m across, a result that is dependent on the impactor size distribution (Fig.~\ref{fig:tau_vs_d_alpha}). In other words, we do not see these older fragments because they have either been removed from the Trojan clouds outright or otherwise converted to smaller - and thus currently unobservable - fragments by collisions. Another potential problem is also related to the rate of mobility due to the Yarkovsky
effect. It would be interesting to evaluate the probability that, in an environment characterised by a low collision rate, 
a 2-km body could still be found in the stability region in spite of it experiencing a dynamical evolution 
driven by Yarkovsky. According to \citet{Cuk.et.al2015}, the dynamical evolution of Mars Trojans, 
taking into account only gravitational perturbations by major planets, is such that the eccentricity can be pumped up to values
sufficient for it to leave the stability region over timescales as short as 1 Gyr. By adding a size-dependent Yarkovsky acceleration,
the residence time can either increase or decrease, depending on the sense of rotation (which determines the direction of
the Yarkovsky-driven drift in semi-major axis). The problem of keeping Eureka in the stable region can be relevant if we admit for 
it a quite old age, as suggested by its uncommon composition. 
\item \emph{Compound scenarios}. In principle, one could conceive of more convoluted models, including an 
interplay of different episodes of rotational fission, cratering or collisional fragmentation that occurred at different epochs. Also the 
interplay of YORP cycles modifying the obliquity angle, and corresponding Yarkovsky acceleration can play a role. Such models appear premature today, in the absence of a much better understanding of the current
inventory and size distribution of Mars Trojans, complete down to smaller sizes, because these parameters have a strong
bearing on any interpretation attempts.
\end{itemize} 

It is certain that constructing a quantitative evolutionary model consistent with YORP dynamics and the observed properties of the 
family would be highly desirable at this stage. On the observational front, completing the family inventory down to 200-300m ($H\sim 20$) objects is sorely needed to develop more credible models of formation of the family but also to add confidence to the orbital correlation found by \citet{Cuk.et.al2015}. Moreover, this could also confirm the similarity to the size distribution of the small MBA population \citep{Gladman.et.al2009,Jacobson.et.al2014}.
Ultimately, however, confirmation of the impact formation model would require {\it in situ} measurements, for instance high-resolution imaging during a flyby to search for evidence of a large-scale cratering event on Eureka's surface \citep{Thomas1999}.

Ground-based photometric investigations aimed at deriving the spin rate and possibly the pole orientation and sense of rotation of Eureka family members would also add much desired information. We note that the unique dynamical environment of the Martian Trojan clouds offers an interesting opportunity to study YORP-driven production of asteroids, likely the dominant source mechanism for small MBAs \citep{Jacobson.et.al2014}. While identifying MBA ``offspring'' is hampered by their high number density in orbital element space \citep{VokrouhlickyNesvorny2008,Pravec.et.al2010}, this is not an issue for the orbitally isolated Eureka family.

Finally, due to the strong dependence of radiation-driven forces on heliocentric distance \citep{Bottke.et.al2006}, the lessons learned from studying this unique population of inner solar system asteroids will apply to long-lived co-orbitals of our own planet, both known and unknown \citep{ChristouAsher2011,Cuk.et.al2012,Dvorak.et.al2012}, to help understand the processes that dominate their origin and evolution.  

\section*{Acknowledgements}
Work by AAC, GB, AC \& AD reported in this paper was supported via a grant (ST/M000834/1) from the UK Science and Technology Facilities Council. 
This publication makes use of data products from the Wide-field Infrared Survey Explorer, which is a joint project of the University of California, Los Angeles, 
and the Jet Propulsion Laboratory/California Institute of Technology, and NEOWISE, which is a project of the Jet Propulsion Laboratory/California Institute of Technology.
AAC gratefully acknowledges support (Short Visit Grant \#6231) from the Gaia Research in European Astronomy Training collaboration within the framework of the European Science Foundation.
Astronomical research at the Armagh Observatory and Planetarium is grant-aided by the Northern Ireland Department for Communities.
\bibliographystyle{elsarticle-harv}
\bibliography{icarus_2016_403}
\clearpage
\protect
\listoffigures

\renewcommand{\baselinestretch}{1.0}
\protect

\newpage
\begin{table}[htb]
\centering
\caption[Mean orbital elements and physical properties of Eureka family asteroids.]{Mean orbital elements and physical properties of Eureka family asteroids.}
\begin{tabular}{cccccc}
\noalign{\smallskip}
\hline \hline
\noalign{\smallskip}
                         &    $D$   &             &    $I$   &   & $\mbox{Diam.}^{\dagger}$ \\
Designation  &   (deg)   &   $e$   &  (deg)   &  $H$  & (km) \\ \hline
\noalign{\smallskip}
(5261) Eureka & 5.63 & 0.0593 & 22.22 & 16.1 & 1.79 \\
(385250) 2001 $\mbox{DH}_{47}$ & 5.90 & 0.0572 & 22.80 & 18.9 & 0.49 \\
(311999) 2007 $\mbox{NS}_{2}$ & 7.40 & 0.0468 & 20.95 &18.1 & 0.71 \\
2011 $\mbox{SC}_{191}$ & 9.52 & 0.0734 & 19.14 &19.3 & 0.41 \\
2011 $\mbox{SL}_{25}$ & 7.97 & 0.0850 & 21.75 & 19.5 & 0.37 \\
2011 $\mbox{UB}_{256}$ & 5.89 & 0.0565 & 22.64 & 20.1 & 0.28 \\
2011 $\mbox{UN}_{63}$ & 7.44 & 0.0512 & 21.60 & 19.7 & 0.34 \\
\noalign{\smallskip} \hline \hline \noalign{\smallskip} 
 \multicolumn{6}{l}{\parbox{100mm}{Orbital elements are from \citet{Cuk.et.al2015}. 
 Absolute magnitudes were retrieved from the Minor Planet Center Database on 26 July 2016. \\
 ${}^{\dagger}$Calculated from $H$ assuming a visible albedo of 0.2.}}
\end{tabular} 
\label{tab:eureka family}
\end{table} 

\clearpage
\begin{table}[htb]
\centering
\caption[Proper semimajor axes and libration amplitudes ($2$$D$) for Eureka family asteroids.]{Proper semimajor axes and libration amplitudes ($2$$D$) for Eureka family asteroids.}
\begin{tabular}{llllllll}
\noalign{\smallskip}
\hline \hline
\noalign{\smallskip}
                   &    (5261)             & (311999)  & (385250)   &   2011  &   2011   &  2011  &   2011 \\ 
                   &    Eureka             & 2007 $\mbox{NS}_{2}$  & 2001 $\mbox{DH}_{47}$   & $\mbox{SC}_{191}$      & $\mbox{SL}_{25}$   & $\mbox{UB}_{256}$  & $\mbox{UN}_{63}$  \\                    
                   \hline \noalign{\smallskip} 
${\Delta \lambda}^{\dagger}$ (${}^{\circ}$) & 11.26 & 14.80 & 11.80 & 19.05 & 15.94 & 11.78 & 14.88 \\
$\Delta \lambda$ (This work) & 11.56  &  14.19 & 10.68 & 14.44 & 11.86 & 13.67 & 17.83 \\
${\Delta a}^{\ddagger}$ (m $\mbox{s}^{-1}$) &  2.39   &  2.94 &  2.21 & 2.99 & 2.46 & 2.83 & 3.69 \\
\noalign{\smallskip} \hline \hline  \noalign{\smallskip} 
\multicolumn{8}{l}{\parbox{167mm}{${}^{\dagger}$ Data from \citet{Cuk.et.al2015}. \\ ${}^{\ddagger}$ Calculated from the row above as $\Delta a = \sqrt{3 \mu_{\rm Mars}} \mbox{ }a_{\rm Mars}$ \citep{Milani1993}.}}
\end{tabular} 
\label{tab:proper}
\end{table} 

\clearpage
\begin{table}[htb]
\centering
\caption[Comparison between analytical model predictions and numerical simulations of the retainment and escape of Trojan fragments.]{Comparison between analytical model predictions and numerical simulations of the retainment and escape of Trojan fragments.}
\begin{tabular}{lllrr}
\hline \hline 
\noalign{\smallskip}
                              &     & \multicolumn{2}{c}{Run \#1: $\rho=2500$ kg $\mbox{m}^{-3}$, $\alpha=2.5$, $D_{\rm LR}=1300$m,}         &           \\ 
                              &      & \multicolumn{2}{c}{ $s_{\rm obs}=70$m, $r=0.8$}          &              \\  \noalign{\smallskip} \hline           \noalign{\smallskip}
                              & \multicolumn{2}{l}{\bf Cellino et al ejection}                                                                 &   \multicolumn{2}{r}{\bf Carruba et al ejection}  \\        
                                                     & \multicolumn{2}{l}{\bf (cratering regime)}                                                                 &   \multicolumn{2}{r}{\bf ($\mathbf{f_{\rm KE}=0.01}$, $\mathbf{Q^{\ast}=10^{3}}$  J $\mbox{\bf kg}^{\bf -1}$)}  \\      \noalign{\smallskip}       
                              &    $N_{\rm Ej}/N_{\rm Total}$           &  $N_{\rm Troj}/N_{\rm Total}$  &    $N_{\rm Ej}/N_{\rm Total}$           &  $N_{\rm Troj}/N_{\rm Total}$   \\  \noalign{\smallskip} \hline  \noalign{\smallskip}
  Model                 &                      0.30                              &                0.70                           &                        0.06                            &                0.94                             \\   \noalign{\smallskip}
    Simulation       &                  $389/1069$                         &           $680/1069$                       &                     $97/1039$                        &            $ 942/1039 $                      \\  \noalign{\smallskip} \hline \noalign{\smallskip}
                              &         &  \multicolumn{2}{c}{Run \#2: $\rho=2500$ kg $\mbox{m}^{-3}$, $\alpha=2.5$, $D_{\rm LR}=3000$m,}    &                \\ 
                              &          &  \multicolumn{2}{c}{ $s_{\rm obs}=100$m, $r=0.5$}      &              \\  \noalign{\smallskip} \hline   \noalign{\smallskip}                        
                              & \multicolumn{2}{l}{{\bf Cellino et al ejection}}                                                                 &   \multicolumn{2}{r}{{\bf Carruba et al ejection}}  \\            
                                                     & \multicolumn{2}{l}{{\bf (catastrophic regime)}}                                                                 &   \multicolumn{2}{r}{\bf ($\mathbf{f_{\rm KE}=0.1}$, $\mathbf{Q^{\ast}=10^{4}}$  J $\mbox{\bf kg}^{\bf -1}$)}  \\  \noalign{\smallskip}           
                              &    $N_{\rm Ej}/N_{\rm Total}$           &  $N_{\rm Troj}/N_{\rm Total}$  &    $N_{\rm Ej}/N_{\rm Total}$           &  $N_{\rm Troj}/N_{\rm Total}$   \\   \noalign{\smallskip} \hline   \noalign{\smallskip}
  Model                 &                   $   0.86  $                            &           $     0.14     $                      &                   $     0.84   $                         &             $   0.16   $                          \\  \noalign{\smallskip}
    Simulation       &                 $ 1251/1429 $                        &       $    178/1429   $                    &                  $   371/415   $                     &        $     44/415    $            \\       \noalign{\smallskip}
\hline
\end{tabular} 
\label{tab:tests}
\end{table} 

\renewcommand{\baselinestretch}{2.0}

\clearpage

\clearpage

\clearpage

\clearpage

\clearpage

\clearpage

\clearpage

\clearpage
\begin{figure}
\vspace{-3cm}
\centering
\includegraphics[width=150mm,angle=0]{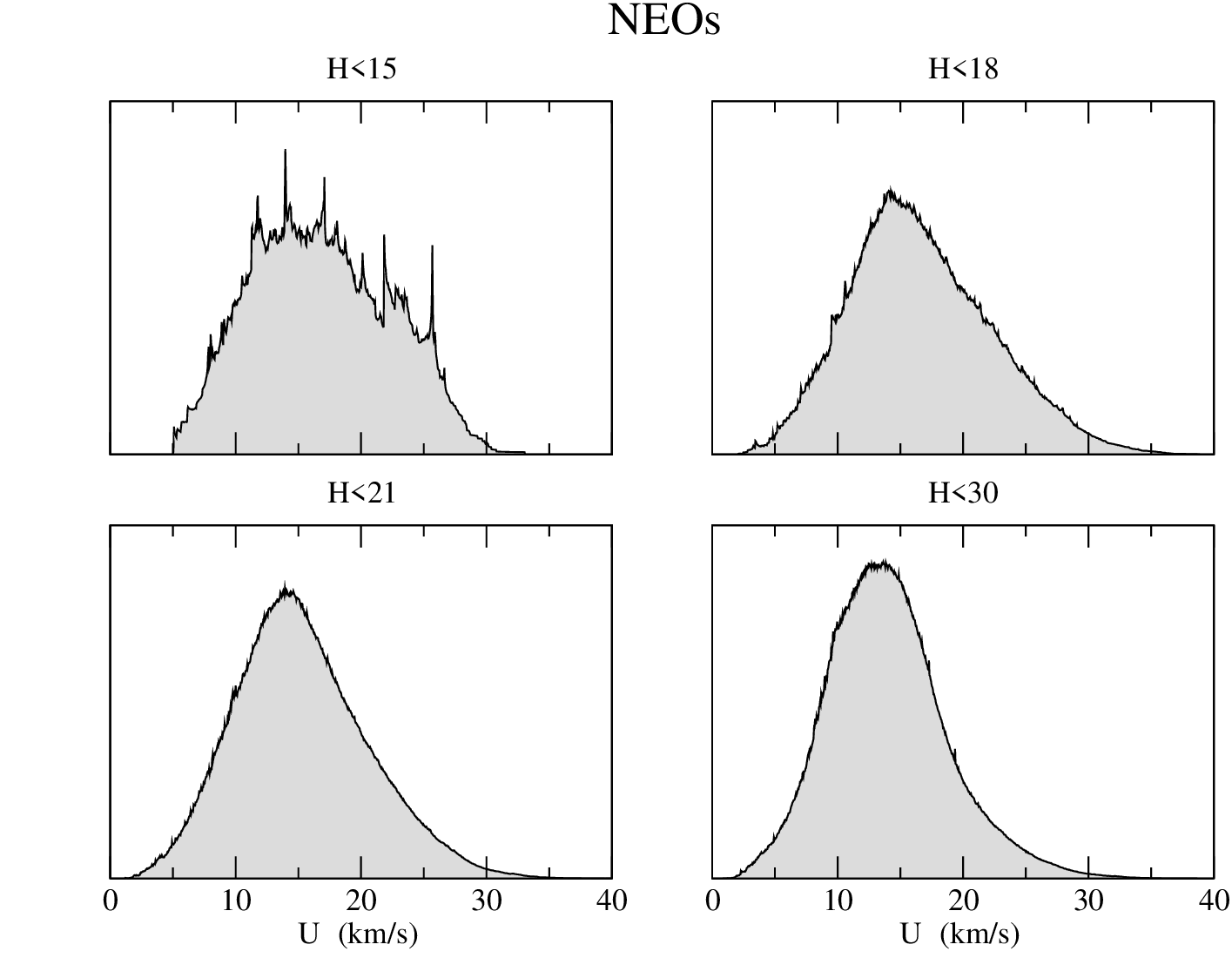}
\caption[As in Fig.~\ref{fig:V_vs_h_mc} but for the Near Earth Object impactor population (NEO).]{}
\label{fig:V_vs_h_neo}
\end{figure}

\clearpage
\begin{figure}
\vspace{-3cm}
\centering
\includegraphics[width=150mm,angle=0]{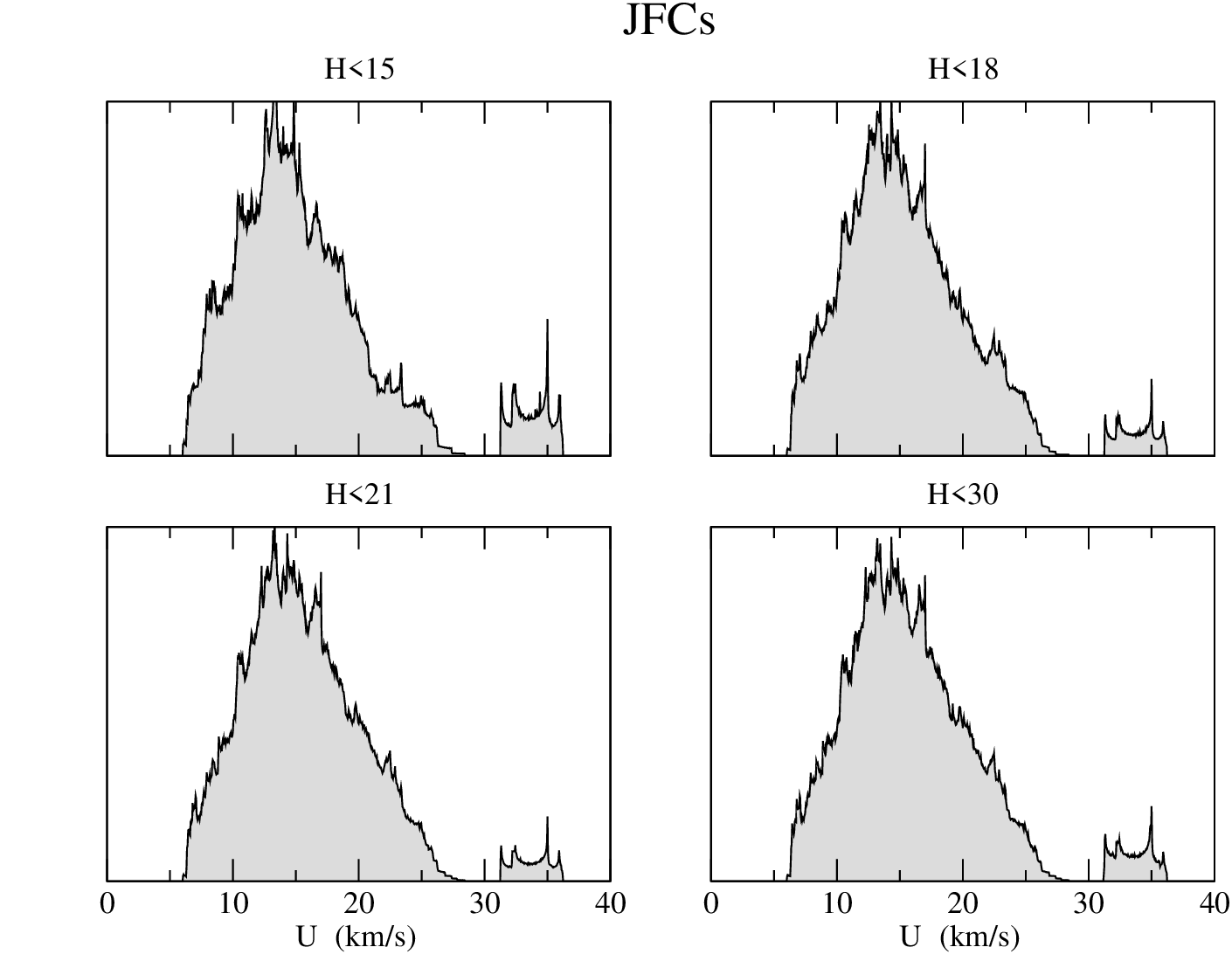}
\caption[As in Fig.~\ref{fig:V_vs_h_mc} but for the Jupiter Family Comet impactor population (JFC).]{}
\label{fig:V_vs_h_jfc}
\end{figure}

\clearpage

\clearpage

\clearpage

\clearpage

\clearpage
\appendix
\section{Derivation of pdfs for Model A}
Combining Eqs.~\ref{eq:V} to \ref{eq:cdf_size} we obtain the joint speed-size pdf
\begin{equation}\label{eq:pdf_v_s_cellino}
p({\rm v},s) = 10^{-0.59}\left(\frac{\alpha}{s_{\rm min}}\right)\left(\frac{s_{\rm LR}}{s_{\rm PB}}\right)^{-0.95} \left(\frac{s_{\rm min}}{s_{\rm PB}}\right)^{\alpha+1} \left(\frac{s}{s_{\rm PB}}\right)^{3/2 - \alpha - 1}
\end{equation} 
that is nonzero within the domain $\left(0, V^{\ast}(s) \right) \times \left(s_{\rm min},s_{\rm LR} \right)$.

For this pdf, the marginal distribution for the speed is
\begin{eqnarray}\label{eq:pdf_v_cellino}
p({\rm v}) &=&10^{-0.59} \left(\frac{\alpha}{\alpha-3/2}\right) \left(\frac{s_{\rm LR}}{s_{\rm PB}}\right)^{-0.95} \nonumber \\
                 &  \times  & \left(\frac{s_{\rm min}}{s_{\rm PB}}\right)^{3/2} \left[1 - \left(s_{\rm min}/s_{\rm U}\right)^{\alpha - 3/2}\right]\mbox{, } 0 \leq {\rm v}<V^{\ast}(s_{\rm min})
\end{eqnarray} 
where $s_{U}=\min\{ s_{\rm LR},S^{\ast}({\rm v})\}$ and the function $S^{\ast}$ is obtained by solving Eq.~\ref{eq:V} for $s$.

For $s_{\rm LR}/s_{\rm PB} \gtrsim 0.8 $, $K^{\prime}$ is independent of the mass ratio $r$ \citep{Cellino.et.al1999}. In that case, expressions (\ref{eq:pdf_v_s_cellino}) and (\ref{eq:pdf_v_cellino}) respectively become
\begin{equation}\label{eq:pdf_v_s_cellino_cratering}
p({\rm v},s) = 10^{0.2} \left(\frac{\alpha}{s_{\rm min}}\right)\left(\frac{s_{\rm min}}{s_{\rm PB}}\right)^{\alpha+1} \left(\frac{s}{s_{\rm PB}}\right)^{3/2 - \alpha - 1}
\end{equation} 
and 
\begin{eqnarray}\label{eq:pdf_v_cellino_cratering}
p({\rm v}) &=& 10^{0.2} \left(\frac{\alpha}{\alpha-3/2}\right) \left(\frac{s_{\rm min}}{s_{\rm PB}}\right)^{3/2} \nonumber \\
                 &  \times &  \left[1 - \left(s_{\rm min}/s_{\rm U}\right)^{\alpha - 3/2}\right]\mbox{, } 0 \leq {\rm v}<V^{\ast}(s_{\rm min})\mbox{.}
\end{eqnarray} 

Introducing the minimum observable fragment size $s_{\rm obs}$, we obtain for the catastrophic case
\begin{equation}\label{eq:pdf_v_s_cellino_obs}
p_{\rm obs}({\rm v},s)= 10^{-0.59}\left(\frac{\alpha}{s_{\rm PB}}\right) \left(\frac{s_{\rm LR}}{s_{\rm PB}}\right)^{-0.95} \left(\frac{s_{\rm obs}}{s_{\rm PB}}\right)^{\alpha} \left(\frac{s}{s_{\rm PB}}\right)^{3/2 - \alpha - 1}
\end{equation}
and
\begin{eqnarray}\label{eq:pdf_v_cellino_obs}
p_{\rm obs}({\rm v})&=&10^{-0.59}\left(\frac{\alpha}{\alpha-3/2}\right) \left(\frac{s_{\rm LR}}{s_{\rm PB}}\right)^{-0.95}\nonumber \\
                      & \times  & \left(\frac{s_{\rm obs}}{s_{\rm PB}}\right)^{3/2} \left[1 - \left(s_{\rm obs}/s_{\rm U}\right)^{\alpha - 3/2}\right] \mbox{, } 0 \leq {\rm v}<V^{\ast}(s_{\rm obs})
\end{eqnarray}
respectively.

The pdf for $\Delta a$ is a branch function. We provide here the full expressions for both the catastrophic:
\begin{eqnarray}
p_{CE99}(\Delta a) & = & 10^{-0.59}\left(\frac{\alpha}{\alpha-\frac{3}{2}}\right) \left(\frac{s_{\rm LR}}{s_{\rm PB}}\right)^{-0.95} \left(\frac{s_{\rm obs}}{s_{\rm PB}}\right)^{3/2}
 \nonumber \\ 
              & \times  &  \left\{\begin{array}{ll}\log \left(\frac{V^{\ast}_{\rm obs}}{|\Delta a|}\right) - 3 \left(\frac{s_{\rm obs}}{10^{0.39} s_{\rm LR}^{0.63} s^{0.37}_{\rm PB}}\right)^{\alpha - \frac{3}{2}} \left[V^{\ast \frac{2 \alpha}{3} - 1}_{\rm obs}  - |\Delta a|^{\frac{2 \alpha}{3} - 1}\right]  / \left(2 \alpha - 3 \right)&  \\
              
 \log \left(\frac{V^{\ast}_{\rm obs}}{|\Delta a|}\right)  - 3 \left(\frac{s_{\rm obs}}{10^{0.39} s_{\rm LR}^{0.63} s^{0.37}_{\rm PB}}\right)^{\alpha - \frac{3}{2}} 
\left[ V^{\ast \frac{2 \alpha}{3} - 1}_{\rm obs} - V^{\ast \frac{2 \alpha}{3} - 1}_{\rm LR} \right] / \left(2 \alpha - 3 \right)
                       \end{array}
               \right. \nonumber \\
             &  &\left.\begin{array}{c}
              + 0   \\
              - \log \left(\frac{V^{\ast}_{\rm LR}}{|\Delta a|} \right)\left(\frac{s_{\rm obs}}{s_{\rm LR}}\right)^{\alpha - \frac{3}{2}} 
               \end{array}
                 \right\}  \begin{array}{l}
                     \mbox{if  $ V^{\ast}_{\rm LR} <|\Delta a | < V^{\ast}_{\rm obs}$} \\
                       \mbox{if $ 0 < |\Delta a| < V^{\ast}_{\rm LR}$}
                         \end{array}
\label{eq:pdf_da_cellino_cat}
\end{eqnarray}

and cratering case:
\begin{eqnarray}
p_{CE99}(\Delta a) & = & 10^{0.2}\left(\frac{\alpha}{\alpha-\frac{3}{2}}\right) \left(\frac{s_{\rm obs}}{s_{\rm PB}}\right)^{3/2}
 \nonumber \\ 
              & \times  &  \left\{\begin{array}{ll}\log \left(\frac{V^{\ast}_{\rm obs}}{|\Delta a|}\right) - 3 \left(\frac{s_{\rm obs}}{0.74 s_{\rm PB}}\right)^{\alpha - \frac{3}{2}} \left[V^{\ast \frac{2 \alpha}{3} - 1}_{\rm obs}  - |\Delta a|^{\frac{2 \alpha}{3} - 1}\right]  / \left(2 \alpha - 3 \right)&  \\
              
 \log \left(\frac{V^{\ast}_{\rm obs}}{|\Delta a|}\right)  - 3 \left(\frac{s_{\rm obs}}{0.74 s_{\rm PB}}\right)^{\alpha - \frac{3}{2}} 
\left[ V^{\ast \frac{2 \alpha}{3} - 1}_{\rm obs} - V^{\ast \frac{2 \alpha}{3} - 1}_{\rm LR} \right] / \left(2 \alpha - 3 \right)
                       \end{array}
               \right. \nonumber \\
             &  &\left.\begin{array}{c}
              + 0   \\
              - \log \left(\frac{V^{\ast}_{\rm LR}}{|\Delta a|} \right)\left(\frac{s_{\rm obs}}{s_{\rm LR}}\right)^{\alpha - \frac{3}{2}} 
               \end{array}
                 \right\}  \begin{array}{l}
                     \mbox{if  $ V^{\ast}_{\rm LR} <|\Delta a | < V^{\ast}_{\rm obs}$} \\
                       \mbox{if $ 0 < |\Delta a| < V^{\ast}_{\rm LR}$}\mbox{.}
                         \end{array}
\label{eq:pdf_da_cellino_cra}
\end{eqnarray}

\section{Formal variance of $\Delta a$ for Model A}
\subsection*{{\rm Catastrophic case:}}
\begin{eqnarray}
V_{CE99}(\Delta a) & = & 10^{-0.59}\left(\frac{\alpha}{\alpha-\frac{3}{2}}\right) \left(\frac{s_{\rm LR}}{s_{\rm PB}}\right)^{-0.95} \left(\frac{s_{\rm obs}}{s_{\rm PB}}\right)^{3/2}   \nonumber \\ 
              & \times  &  \left[\frac{2}{9} \left( 1+3 \log{\frac{V^{\ast}_{\rm obs}}{V^{\ast}_{\rm LR}}}\right) V^{\ast 3}_{\rm LR}  \right. \nonumber \\
              &              & \left.  - s^{\alpha - 3/2}_{\rm obs} \left(2.5 s^{0.63}_{\rm LR} s^{0.37}_{\rm PB} \right)^{- \left(2 \alpha / 3 - 1\right)} V^{\ast 2 \alpha / 3 - 1}_{\rm obs} V^{\ast 3}_{\rm LR} /\left( 2 \alpha / 3 - 1\right) \right.  \nonumber \\ 
              &  & \left. - \frac{3}{3 + \alpha}  V^{\ast 2 \alpha / 3 + 2}_{\rm LR} - \frac{2}{9}V^{\ast 3}_{\rm LR}  \left(\frac{s_{\rm obs}}{s_{\rm LR}}\right)^{\alpha - 3/2} \right. \nonumber \\ 
              &  & \left. +  \left( V^{\ast 3}_{\rm obs} - V^{\ast 3}_{\rm LR} + 3 V^{\ast 3}_{\rm LR} \log{\frac{V^{\ast}_{\rm obs}}{V^{\ast}_{\rm LR}}} \right) \right. \nonumber \\
              &  & \left. - s^{\alpha- 3/2}_{\rm obs} \left(2.5 s^{0.63}_{\rm LR} s^{0.37}_{\rm PB} \right)^{- \left( \alpha - 3/2\right)} {\left[6 \left(3 + \alpha \right)  \left(2 \alpha /3  - 1 \right) V^{\ast}_{\rm obs}\right]}^{-1} \right. \nonumber \\
              &   & \left. \times \left(-3  V^{\ast 2 \alpha/3 + 3}_{\rm obs} + 2  \alpha  V^{\ast  2 \alpha/3 + 3}_{\rm obs}  - 6  V^{\ast 2 \alpha/3}_{\rm obs} V^{\ast 3}_{\rm LR}  \right.  \right. \nonumber \\
               &   & \left. \left. - 2 \alpha V^{\ast 2 \alpha/3}_{\rm obs} V^{\ast 3}_{\rm LR}  + 9  V^{\ast}_{\rm obs} V^{\ast 2 \alpha/3 + 2}_{\rm LR} \right)              
               \right]  \mbox{.} 
\label{eq:var_da_cellino_cat}
\end{eqnarray}
\subsection*{{\rm Cratering case:}}
\begin{eqnarray}
V_{CE99}(\Delta a) & = & 10^{0.2}\left(\frac{\alpha}{\alpha-\frac{3}{2}}\right) \left(\frac{s_{\rm obs}}{s_{\rm PB}}\right)^{3/2}  \nonumber \\ 
              & \times  &  \left[\frac{2}{9} \left( 1+3 \log{\frac{V^{\ast}_{\rm obs}}{V^{\ast}_{\rm LR}}}\right) V^{\ast 3}_{\rm LR}  \right. \nonumber \\
              &              & \left.  - 2 s^{\alpha - 3/2}_{\rm obs} \left(0.74 s_{\rm PB} \right)^{- \left(\alpha - 3/2\right)} V^{\ast 2 \alpha / 3 - 1}_{\rm obs} V^{\ast 3}_{\rm LR} /\left( 2 \alpha - 3\right) \right.  \nonumber \\ 
              &  & \left. - \frac{3}{3 + \alpha}  V^{\ast 2 \alpha / 3 + 2}_{\rm LR} - \frac{2}{9}V^{\ast 3}_{\rm LR}  \left(\frac{s_{\rm obs}}{s_{\rm LR}}\right)^{\alpha - 3/2} \right. \nonumber \\ 
              &  & \left. +  \frac{1}{9}\left( V^{\ast 3}_{\rm obs} - V^{\ast 3}_{\rm LR} + 3 V^{\ast 3}_{\rm LR} \log{\frac{V^{\ast}_{\rm obs}}{V^{\ast}_{\rm LR}}} \right) \right. \nonumber \\
              &  & \left. - s^{\alpha- 3/2}_{\rm obs} \left(0.74 s_{\rm PB} \right)^{- \left( \alpha - 3/2\right)} {\left[6 \left(3 + \alpha \right)  \left(2 \alpha /3  - 1 \right) V^{\ast}_{\rm obs}\right]}^{-1} \right. \nonumber \\
              &   & \left. \times \left(-3  V^{\ast 2 \alpha/3 + 3}_{\rm obs} + 2  \alpha  V^{\ast  2 \alpha/3 + 3}_{\rm obs}  - 6  V^{\ast 2 \alpha/3}_{\rm obs} V^{\ast 3}_{\rm LR}  \right.  \right. \nonumber \\
               &   & \left. \left. - 2 \alpha V^{\ast 2 \alpha/3}_{\rm obs} V^{\ast 3}_{\rm LR}  + 9  V^{\ast}_{\rm obs} V^{\ast 2 \alpha/3 + 2}_{\rm LR} \right)              
               \right]  \mbox{.} 
\label{eq:var_da_cellino_cra}
\end{eqnarray}
\end{document}